\documentclass[11pt,reqno]{amsart}

\usepackage{epsfig}
\usepackage{amsmath,amssymb}

\usepackage{amstext}
\usepackage{amscd}
\usepackage{latexsym}

\usepackage{amsfonts}

\numberwithin{equation}{section}

\newcommand{\wt}{\widetilde}
\newcommand{\wh}{\widehat}
\newcommand{\PP}{{\mathbb P}}
\newcommand{\R}{{\mathbb R}}
\newcommand{\C}{{\mathbb C}}

\newcommand{\Z}{{\mathbb Z}}

\newcommand{\CP}{\C\PP}
\newcommand{\RP}{\R\PP}

\renewcommand{\phi}{\varphi}

\newcommand{\dcal}{\mathcal{D}}

\newcommand{\al}{\alpha}
\newcommand{\be}{\beta}
\newcommand{\ga}{\gamma}

\newcommand{\La}{\Lambda}
\newcommand{\la}{\lambda}
\newcommand{\ep}{\varepsilon}
\newcommand{\eps}{\epsilon}
\newcommand{\de}{\delta}
\newcommand{\De}{\Delta}
\newcommand{\f}{\varphi}
\newcommand{\om}{\omega}
\newcommand{\sg}{\sigma}
\newcommand{\z}{\zeta}

\newtheorem{theo}{{\sc Theorem}}[section]
\newtheorem{theorem}{{\sc Theorem}}[section]

\newtheorem{lem}[theo]{{\sc Lemma}}
\newtheorem{lemma}[theo]{{\sc Lemma}}
\newtheorem{prop}[theo]{{\sc Proposition}}

\newenvironment{rem}{\medskip\noindent{\it Remark:\/} }{\medskip}
\newenvironment{rems}{\medskip\noindent{\it Remarks:\/} }{\medskip}

 2
 3
 4
 2

\renewcommand{\(}{\left(}
\renewcommand{\)}{\right)}

\newcommand{ \E}{{\mathbf E}\,}

\begin{document}

\title[Non-Gaussian Random Polynomials]
{Correlations between Zeros
of Non-Gaussian Random Polynomials}

\author{ Pavel M. Bleher}

\address{Department of Mathematical Sciences\\
               Indiana University-Purdue University Indianapolis\\
               402 N. Blackford Street\\
               Indianapolis, IN 46202, USA}

\email{bleher@math.iupui.edu.}

\author{Xiaojun Di}

\address{Department of Mathematical Sciences\\
               Indiana University-Purdue University Indianapolis\\
               402 N. Blackford Street\\
               Indianapolis, IN 46202, USA}


\date{\today}

\vskip 5mm

\begin{abstract} The existence of the scaling limit and its
  universality, for correlations between zeros of  {\it
    Gaussian}   random polynomials, or more generally, {\it Gaussian}
  random sections  
  of powers of a line bundle over a compact manifold has been 
  proved in a great generality in the works \cite{BBL2}, \cite {Ha},
  \cite {BD}, \cite{BSZ1}-\cite{BSZ4}, and others. In the present work we
  prove the existence of the scaling limit for a class of {\it non-Gaussian}
  random polynomials. Our main result is that away from the origin the 
  scaling limit exists and is universal, so that it does not depend on
  the distribution of the coefficients. At the origin the scaling
  limit is not universal, and  we find a crossover from the
  nonuniversal asymptotics of the density of the probability distribution
  of zeros at the origin to the universal one away
  from the   origin.  
\end{abstract}

\maketitle

\section{Introduction}\label{intro}

Random  
polynomials, or more generally, linear combinations of functions
with random coefficients serve as a basic model for eigenfunctions of
chaotic quantum systems, see \cite{BBL1}, \cite{BBL2}, \cite {LS},
\cite {Ha}, \cite {NV}, \cite {Leb}, \cite{Be}, and others. The
geometric 
structure of random polynomials is, therefore, of significant interest 
for applications to quantum chaos. The basic questions are
distribution and correlations between zeros of random polynomials,
their 
critical points, distribution of values, nodal lines and surfaces,
etc., see \cite{BBL1}, \cite{BBL2}, \cite {LS},
\cite {Ha}, \cite {NV}, \cite {Leb}, \cite{Be},
\cite{BeD1}, \cite {BeD2}, \cite {MSG}, and others.
The principal problem concerns the asymptotic behavior of  
typical, in the probabilistic sense, geometric structures as the
degree of the polynomial, or the number of terms in a linear
combination of functions, goes to infinity.
The distribution of zeros of random polynomials is a classical
question in probability theory and we refer to the book \cite {BhS}
and to the paper \cite{EK} for 
many earlier results in this area. The convergence of the
distribution of zeros and their correlation functions in the scaling
limit  has been  proved for {\it Gaussian} ensembles of random linear
combinations of functions in a great generality, see \cite {BBL1},
\cite {BBL2}, 
\cite{Ha}, \cite{BD}, \cite {MBFMA}, \cite{BSZ1}-\cite{BSZ4}, \cite
{BR}, \cite {SZ1}-\cite{SZ3}, and others. In the present paper we
address the same question for {\it non-Gaussian} ensembles. We will
mostly consider the non-Gaussian SO(2) ensemble of random polynomials
but our approach is quite general.
In the last section we will discuss extensions to multivariate and
complex ensembles of random polynomials.

Consider the real random polynomial of the form
\begin{equation}\label{POLY}
f_n(x) = \sum_{k=0}^n \sqrt{\binom{n}{k}}\,c_kx^k,
\qquad \binom{n}{k}=\frac{n!}{k!(n-k)!}\,,
\end{equation}
where $c_k$ are independent identically distributed random variables
such that
\begin{equation} \label{CK}
{\E}c_k = 0, \qquad {\E}c_k^2 = 1.
\end{equation}
When $c_k=N(0,1)$, the standard Gaussian random variable, the distribution
of real zeros of $f_n(x)$ has two remarkable properties:
\begin{enumerate} 
\item The mathematical expectation of the number of real
zeros of $f_n(x)$ is equal to $\sqrt{n}$,
\begin{equation} \label{so2a}
\E \,\#\{ k:\, f_n(x_k)=0\}=\sqrt{n};
\end{equation}
\item The normalized
distribution of real zeros on the real line is the Cauchy distribution,
\begin{equation} \label{so2b}
\frac{1}{\sqrt n}\, \E\, \#\{ k:\, f_n(x_k)=0,\;a\le x_k \le b\}=
\int_a^b \frac{1}{\pi (x^2+1)}dx,
\end{equation}
\end{enumerate}
see \cite{EK}, \cite{BD}.
The both properties are exact for any finite $n$. 
The second property can be reformulated as follows. Set
\begin{equation} \label{stereo}
\theta_k=\arctan x_k,
\end{equation}  
so that $\theta_k$ is the stereographic projection of $x_k$.
Then $\theta_k$'s are uniformly distributed on the circle 
$$
-\frac{\pi}{2}\le \theta\le \frac{\pi}{2},\quad -\frac{\pi}{2}\equiv
\frac{\pi}{2}\,.
$$
For $\theta_k$'s a stronger statement is valid: if $K_{nm}(s_1,\ldots,s_m)$
is the $m$-point correlation function of $\theta_k$'s, then for any $a$,
\begin{equation} \label{stereo1}
K_{nm}(s_1+a,\ldots,s_m+a)=K_{nm}(s_1,\ldots,s_m),
\end{equation}
the SO(2)-invariance \cite{BD}. As shown in \cite{BD}, as $n\to\infty$,
there exists the scaling limit of correlation functions,
\begin{equation} \label{stereo2}
\lim_{n\to\infty}\frac{1}{n^{m/2}}K_{nm}(s+\frac{\tau_1}{\sqrt n},\ldots,s+
\frac{\tau_m}{\sqrt n})=K_m(\tau_1,\ldots,\tau_m),
\end{equation}
with expicit formulae for the limiting correlation functions.

In the present work we will be interested in an extension of these
results to non-Gaussian $c_k$'s.  
We will assume that the probability distribution of $c_k$ is
absolutely continuous with respect to the Lebesgue measure, 
with a continuous density function $r(t)$. Conditions
on $r(t)$ will be formulated below. In sections
\ref{dist1}-\ref{dist3} below 
we will show the universality of the Cauchy distribution as a limiting 
distribution of real zeros away from the origin. In sections
\ref{scaling1}-\ref{scaling2} we will derive a 
nonuniversal scaling behavior of the 
distribution of real zeros near the origin. In sections
\ref{uni-corr1}-\ref{uni-corr2} we will formulate and prove the
universality of 
limiting correlation functions away from the origin. In the concluding section
\ref{conclusion} we will  discuss extensions of our results
to  multivariate random polynomials. 

\section{Distribution of Real Zeros}\label{dist1}

We will be first interested in the distribution of real 
zeros of $f_n(x)$, and
our calculations will be based on the Kac-Rice formula. The Kac-Rice 
formula \cite{Kac}, \cite{Ric} expresses the density $p_n(x)$ of
the distribution of real zeros of the polynomial $f_n(x)$ as
\begin{equation}\label{DENSITY}
p_n(x)=\int_{-\infty}^\infty |\eta|D_n(0,\eta;x)\,d\eta,
\end{equation}
where $D_n(\xi,\eta;x)$ is the joint distribution density
of $f_n(x)$, $f'_n(x)$. Since $f_n(0)=c_0$, $f_n'(0)=\sqrt{n}\,c_1$,
we obtain that
\begin{equation}\label{pn0}
p_n(0)=\sqrt{n}\,r(0)\int_{-\infty}^\infty |t|\,r(t)\,dt\,.
\end{equation}
In particular, if $c_k=N(0,1)$,
\begin{equation}\label{pn0G}
p_n(0)=\frac{\sqrt{n}}{\pi}\,.
\end{equation}
Assume now that $x\not=0$.

From (\ref{CK}), 
\begin{equation}
{\E}  f_n(x)=0, \quad
{\E}  f^\prime_n(x)=0,
\end{equation}
 and
\begin{equation}
\begin{aligned}
{\E}  f_n^2(x)&=(1+x^2)^n\equiv \sg^2_n(x)\,,  \\
{\E}  f_n(x)f_n^{\prime}(x)&=nx(1+x^2)^{n-1}\,, \\
{\E}  (f_n^{\prime}(x))^2&= n(1+nx^2)(1+x^2)^{n-2}
\equiv \z^2_n(x)\,.
\end{aligned}
\end{equation}
To study the limit as $n \to \infty$, it is useful to
rescale 
$f_n(x)$, $f^\prime_n(x)$.  Let
\begin{equation}\label{gn}
g_n(x)\equiv \frac{f_n(x)}{\sg_n(x)}
=\sum_{k=0}^n\mu_k(x) c_k,\quad \wt g_n(x)\equiv \frac{f'_n(x)}{\z_n(x)}
=\sum_{k=0}^n\nu_k(x) c_k\,.
\end{equation}
where
\begin{equation}\label{muk}
\mu_k(x)= \frac{x^k}  {\sg_n(x)}\binom {n}{k}^{1/2}, 
\qquad \nu_k(x) = \frac{kx^{k-1}}{\z_n(x)}\binom{n}{k}^{1/2}, \qquad k
= 0, 1, \dots, n, 
\end{equation}
are the weights.
Let $\wt D_n(\xi,\eta;x)$ be the
the joint distribution density of
$g_n(x)$ and $\wt g_n(x))$.
Then
\[
D_n(\xi,\eta;x)=\frac{1}{\sg_n(x)\z_n(x)}\,
\wt D_n\(\frac{\xi}{\sg_n(x)},\frac{\eta}{\z_n(x)};x\),
\]
hence equation (\ref{DENSITY}) reduces to
\begin{equation}\label{MDENSITY}
p_n(x)=\frac{\z_n(x)}{\sg_n(x)}\int_{-\infty}^\infty |\eta|\wt
D_n(0,\eta;x)\,d\eta. 
\end{equation}
From (\ref{muk}),
\begin{equation} \label{MU_NU}
\mu_k(x)=
\frac{x^k}{(1+x^2)^{n/2}}\binom {n}{k  }^{1/2}\,,\quad
\nu_k(x)=\mu_k(x)\frac{k(1+x^2)}{\sqrt n\,x\sqrt{1+nx^2}}\,,
\end{equation}
and
\begin{equation} \label{MU_NU_R}
\sum_{k=0}^n\mu_k(x)^2=1,\quad \sum_{k=0}^n\nu_k(x)^2=1,\quad
\sum_{k=0}^n\mu_k(x)\nu_k(x)=\frac{x\sqrt n}{\sqrt{1+nx^2}}.
\end{equation}
It is convenient to orthogonalize the pair $g_n(x)$,
$\tilde{g}_n(x)$. To that end  define 
\begin{equation}\label{hn}
h_n(x) \equiv \frac{\tilde{g}_n(x) - (\nu(x),\mu(x))g_n(x)}{\tau_n(x)}
= 
\sum_{k=0}^n\lambda_k(x)c_k, 
\end{equation}
where
\[
\mu(x)=(\mu_0(x),\ldots,\mu_n(x)),\quad
\nu(x)=(\nu_0(x),\ldots,\nu_n(x)),\quad
 (\nu(x),\mu(x))=\sum_{k=0}^n\nu_k(x)\mu_k(x), 
\]
and
\begin{equation} \label{LA}
\begin{aligned} 
\la(x)&=(\la_0(x),\ldots,\la_n(x))
=\frac{\nu(x)-(\nu(x),\mu(x))\mu(x)}{\tau_n(x)}\,;\\
\tau_n(x)&=\|\nu(x)-(\nu(x),\mu(x))\mu(x)\|
=\(\sum_{k=0}^n[\nu_k(x)-(\nu(x),\mu(x))\mu_k(x)]^2\)^{1/2}.
\end{aligned}
\end{equation}
Observe that by (\ref{MU_NU_R}) and (\ref{LA}),
\begin{equation} \label{MU_LA_R}
\sum_{k=0}^n\mu_k(x)^2=1,\quad \sum_{k=0}^n\la_k(x)^2=1,\quad
\sum_{k=0}^n\la_k(x)\mu_k(x)=0. 
\end{equation}
From (\ref{MU_NU}),
\begin{equation}\label{NUK}
\nu_k(x)=\mu_k(x)\frac{u(1+x^2)\sqrt
  n}{x\sqrt{1+nx^2}} 
=\mu_k(x)\frac{u\sqrt n\,x}{u_0\sqrt{1+nx^2}}\,,
\quad u\equiv\frac{k}{n}\,,
\quad u_0\equiv \frac{x^2}{1+x^2}\,.
\end{equation}
hence
\begin{equation} \label{LAK}
\la_k(x)=
\frac{\mu_k(x) \sqrt n \,(u-u_0)(1+x^2)}{x}\,.
\end{equation}
Let $\wh D_n(\xi,\eta;x)$ be the joint distribution density of
$g_n(x)$ and $h_n(x)$. Then
\[
\wt D_n(\xi,\eta;x)
=\frac{1}{\tau_n(x)}\wh
D_n\(\xi,\frac{\eta-(\nu(x),\mu(x))\xi}{\tau_n(x)};x\), 
\]
hence equation (\ref{MDENSITY}) reduces to
\begin{equation} \label{MMDENSITY}
p_n(x)=\frac{\z_n(x)\tau_n(x)}{\sg_n(x)}
\int_{-\infty}^\infty |\eta|\wh D_n(0,\eta;x)\,d\eta.
\end{equation}
From (\ref{MU_NU_R}) and (\ref{LA}),
\begin{equation}\label{numu}
(\nu(x),\mu(x))=\frac{\sqrt{n}\,x}{\sqrt{1+nx^2}},
\quad \tau_n(x)^2=1-(\nu(x),\mu(x))^2=\frac{1}{1+nx^2},
\end{equation}
Hence (\ref{MMDENSITY}) implies that
\begin{equation} \label{RDENSITY}
p_n(x)=\frac{\sqrt n}{(1+x^2)}
\int_{-\infty}^\infty |\eta|\wh D_n(0,\eta;x)\,d\eta.
\end{equation} 
By (\ref{CK}) and (\ref{MU_LA_R}), for all real $x$,
\begin{equation}\label{Egh}
\E g_n(x)=\E h_n(x)=0,\quad \E g_n(x)^2=\E h_n(x)^2=1,
\quad \E g_n(x)h_n(x)=0.
\end{equation}
If $c_k$'s are Gaussian, then $g_n(x)$ and $h_n(x)$ are
Gaussian as well and equation (\ref{RDENSITY}) reduces to
\begin{equation}\label{Gaussp}
p_n(x)=\frac{\sqrt n}{\pi(1+x^2)}\,.
\end{equation}

Let us calculate the asymptotics of the weights $\mu_k(x)$ as $n\to\infty$. 
By the Stirling formula,
\[
\binom {n}{k  }=\frac{1}{\sqrt{2\pi nu(1-u)}}
e^{-n\Theta(u)}\left(1+O(\ep_k)\right),
\quad u=\frac{k}{n}\,,
\]
where
\begin{equation}
\Theta(u)\equiv u\ln u+(1-u)\ln(1-u),
\quad \ep_k\equiv (k+1)^{-1}+(n+1-k)^{-1}.
\end{equation}
Therefore,
\begin{equation} \label{MU_ASYM}
\mu_k(x)^2=\frac{1}{\sqrt{2\pi
    nu(1-u)}}e^{-n\Theta(u;x)}\left(1+O(\ep_k)\right), 
\end{equation}
where
\begin{equation}
\Theta(u;x)\equiv u\ln u+(1-u)\ln(1-u)+\ln(1+x^2)-u\ln x^2.
\end{equation}
The minimum of $\Theta(u;x)$ in $u$ is attained at
\begin{equation}
u_0\equiv \frac{x^2}{1+x^2}\,,
\end{equation}
and
\begin{equation}\label{u0}
\Theta(u_0;x)=0,\qquad \Theta''(u_0;x)=\frac{1}{u_0(1-u_0)}=
\frac{(1+x^2)^2}{x^2}>0
\end{equation}
Observe that
\begin{equation}
\Theta''(u;x)=\frac{1}{u(1-u)}>0\,,
\end{equation}
hence $\Theta(u;x)$ is a convex function on the interval $0<u<1$.
Therefore, we obtain from (\ref{MU_ASYM})
that there exists $C>0$ such that
\begin{equation}\label{max}
\max_{0\le k\le n}|\mu_k(x)|\le \frac{C}{[n u_0(1-u_0)]^{1/4}}
=\frac{C(1+x^2)^{1/2}}{n^{1/4}|x|^{1/2}}. 
\end{equation}
By (\ref{LAK}), (\ref{MU_ASYM}) and (\ref{u0}),
\begin{equation} \label{LAKa}
\la_k(x)^2=
 n \Theta''(u_0;x)\,(u-u_0)^2\,\frac{1}{\sqrt{2\pi
    nu(1-u)}}e^{-n\Theta(u;x)}\left(1+O(\ep_k)\right).
\end{equation}
Simple estimates show that, similarly to (\ref{max}),  there exists
$C>0$ such that
\begin{equation}\label{max1}
\max_{0\le k\le n}|\la_k(x)|\le\frac{C(1+x^2)^{1/2}}{n^{1/4}|x|^{1/2}}. 
\end{equation}
The main results of this section are summarized as follows.

\begin{prop} \label{dens}
For $x\not=0$, the density $p_n(x)$ of the zeros distribution 
of random polynomial (\ref{POLY}) is given by 
formula (\ref{RDENSITY}), where $\wh D_n(\xi,\eta;x)$ is the
joint distribution density of the random variables
\begin{equation}\label{prop1}
g_n(x)=\sum_{k=0}^n\mu_k(x)\,c_k\,,\qquad
h_n(x)=\sum_{k=0}^n\la_k(x)\,c_k\,,
\end{equation}
and
\begin{equation}\label{prop2}
\begin{aligned}
\mu_k(x)&= \frac{x^k}  {(1+x^2)^{n/2}}\binom {n}{k}^{1/2}\,,
\quad
\la_k(x)= \frac{\mu_k(x) \sqrt n \,(u-u_0)(1+x^2)}{x}\,;\\
u&=\frac{k}{n}\,,\quad  
u_0= \frac{x^2}{1+x^2}\,.
\end{aligned}
\end{equation}
For every $x\not=0$, the vectors $\mu(x)=(\mu_0(x),\ldots,\mu_n(x))$ and
$\la(x)=(\la_0(x),\ldots,\la_n(x))$ are orthonormal; cf.
(\ref{MU_LA_R}) and (\ref{Egh}). In addition, there exists $C>0$ such
that for all $x\not=0$,
\begin{equation}\label{MU_NU_LA}
\max_{0\le k\le
  n}|\mu_k(x)|\le\frac{C(1+x^2)^{1/2}}{n^{1/4}|x|^{1/2}}\,,
\qquad
\max_{0\le k\le
  n}|\la_k(x)|\le\frac{C(1+x^2)^{1/2}}{n^{1/4}|x|^{1/2}}\,. 
\end{equation}
For $x=0$, the density $p_n(x)$
is given by formula (\ref{pn0}).
\end{prop}

\section{Universality of the Limiting Distribution
 of Real Zeros}\label{dist2}

Let  $\phi(s)$ be the characteristic function of $c_k$, 
\begin{equation}\label{char1}
\phi(s) = \int_{-\infty}^\infty r(t) e^{its}dt,
\end{equation}
Observe that by (\ref{CK})
\begin{equation}\label{phi}
\phi(0) = 1, \qquad \phi^{\prime}(0) = 0, 
\qquad \phi^{\prime\prime}(0) = -1; \qquad |\phi(s)|\le 1,
\quad s\in\R.
\end{equation}
We will assume that  
$\f(s)$ satisfies the following estimate:
for some $a,\,q>0$,
\begin{equation}\label{C1}
|\f(s)|\le\frac{1}{(1+a s^2)^q}\,,\quad s\in\R.
\end{equation}
In addition, we will assume that $\f(s)$ is a three times
differentiable function and there exist $c_2,c_3>0$
such that
\begin{equation}\label{C2}
\sup_{-\infty<s<\infty}\left|\frac{d^j\f(s)}{ds^j}\right|\le c_j
\quad j=2,3\,.
\end{equation}
Since $\f'(0)=0$, this implies that for real $s$,
\begin{equation}\label{C3}
\left|\frac{d\f(s)}{ds}\right|\le c_2|s|\,.
\end{equation}
Conditions (\ref{C1}), (\ref{C2}) are fulfilled for any density of the form
\begin{equation}\label{C4}
r(t)=e^{-V(t)},
\end{equation}
where $V(t)$ is a polynomial of even degree with a positive 
leading coefficient,
such that
\begin{equation}\label{C5}
\int_{-\infty}^\infty te^{-V(t)}dt=0,
\quad
\int_{-\infty}^\infty t^2e^{-V(t)}dt=1.
\end{equation}
More generally, introduce the following class 
of densities.

{\it Class $\dcal_n$ of densities.} A probability density function
$r(t)$ belongs to the class $\dcal_n$, if
\begin{equation}\label{C5a}
\int_{-\infty}^\infty tr(t)dt=0,
\quad
\int_{-\infty}^\infty t^2r(t)dt=1,
\end{equation} 
$r(t)$ is $C^2$-smooth, and
for any $j= 0,\dots,n$ there exists $C_j>0$ such that
for all $t\in\R$,
\begin{equation}\label{C6}
|r(t)|+|r''(t)|\le \frac{C_j}{1+|t|^j}\,.
\end{equation}
We will denote $\dcal_{\infty}=\cap_{n=0}^{\infty} \dcal_n$.

Conditions (\ref{C1}), (\ref{C2}) are fulfilled for any density
from the class $\dcal_5$.

\begin{theorem}\label{univer}
Let $f_n(x)$ be a random polynomial of degree $n$, as defined in
{\rm{(\ref{POLY})}}. Let $p_n(x)$  
be the distribution density function of real zeros of $f_n(x)$, and
let $\phi(s)$ be 
the characteristic function of $c_k$. If $\phi(s)$ satisfies conditions
(\ref{C1}), (\ref{C2}), then for all $\de>0$,
\begin{equation}\label{uni1}
\lim_{n\to\infty}\frac{p_n(x)}{\sqrt{n}} =\frac{1}{\pi (1+x^2)}\,,
\end{equation}
uniformly for all $x$ such that $\de^{-1}\ge |x|\ge\de$.
This means that
 the normalized distribution density function of real zeros of $f_n(x)$ has 
a universal limit of the Cauchy distribution if $x\not=0$.
\end{theorem}

\begin{rem}
Observe that by (\ref{pn0}),
\begin{equation}\label{uni2}
\frac{p_n(0)}{\sqrt{n}} =r(0)\int_{-\infty}^\infty |t|\,r(t)\,dt\,.
\end{equation}
This shows that at $x=0$ the universal limit (\ref{uni1})
does not hold in general. It is possible
to derive a scaling formula
for $p_n(x)$ in a vicinity of $x=0$, which interpolates between
(\ref{uni2}) and (\ref{uni1}); see section \ref{scaling1} below.
\end{rem}

We will prove Theorem \ref{univer} in the next section.  In fact,
we will prove that for any $\de>0$ and any $\eps>0$ there exists
$C_{\eps}>0$ such that for all $x$ such that $\de<|x|<\de^{-1}$,
\begin{equation}\label{uni2a}
\left|\frac{p_n(x)}{\sqrt{n}} -\frac{1}{\pi (1+x^2)}\right|\le 
C_{\eps}n^{-\frac{1}{12}+\eps}.
\end{equation}
This estimates the rate of convergence of $\frac{p_n(x)}{\sqrt{n}}$
to the Cauchy distribution density, $\frac{1}{\pi (1+x^2)}$.
In subsequent sections we will prove the convergence of correlation
functions. We will assume a stronger condition on the characteristic
function $\phi(s)$: it is $C^\infty$ smooth and for any $j\ge 2$
there is $c_j>0$ such that 
\begin{equation}\label{uni2b}
\left|\frac{d^j\phi(s)}{ds^j}\right|\le c_j
\end{equation}
Under the stronger condition, it will follow the better rate
of convergence for the correlation functions, and, in particular,
for the density,
\begin{equation}\label{uni2c}
\left|\frac{p_n(x)}{\sqrt{n}} -\frac{1}{\pi (1+x^2)}\right|\le 
C_{\eps}n^{-\frac{1}{4}+\eps}.
\end{equation}

\section{Proof of Theorem \ref{univer}}\label{dist3}

Let $\wh D_n(\xi,\eta;x)$ be the
joint distribution density of $g_n(x)$, $h_n(x)$ and  
$\Phi_n(\ga)=\Phi_n(\ga;x)$,
the corresponding characteristic function,
\begin{equation} \label{pro1}
\Phi_n(\ga) = \int_{-\infty}^\infty  \int_{-\infty}^\infty 
\wh D_n(\xi,\eta;x)e^{i\al\xi + i\be\eta}d\xi d\eta
\end{equation}
where $\ga=(\al,\be)$. Our strategy will be to prove that
$\wh D_n(\xi,\eta;x)$ converges, in an appropriate sense,
to the Gaussian density $\frac{1}{2\pi}e^{-\frac{1}{2}(\xi^2+\eta^2)}$.
This will be a Lindeberg type local central limit theorem for 
vector random variables, with an additional
estimate of the tail of the density  $\wh D_n(\xi,\eta;x)$.
First we will prove that the characteristic function $\Phi_n(\ga)$
converges to $e^{-\frac{1}{2}|\ga|^2}$.
From (\ref{gn}), (\ref{hn}) we have that
\begin{equation} \label{pro2}
\Phi_n(\ga)
= \prod_{k=0}^n \phi(\om_k)\,,
\end{equation}
where $\f$ is the characteristic function of $c_k$ and
\begin{equation}\label{pro3}
\om_k= \mu_k(x)\al+ \la_k(x)\be.
\end{equation}

\begin{lemma}\label{univ1}
If $\f(s)$ satisfies (\ref{C1}), then for any $L>0$
there exist $a_0>0$ and $N_0>0$ such that for all $n\ge N_0$,
\begin{equation}\label{pro4}
|\Phi_n(\ga)|\le\frac{1}{(1+a_0 |\ga|^2)^L}\,.
\end{equation}
\end{lemma}

\begin{proof} From (\ref{pro2}) and (\ref{C1}),
\begin{equation}\label{pro5}
|\Phi_n(\ga)|\le \prod_{k=0}^n \frac{1}{(1+a \om_k^2)^q}\,.
\end{equation}
We have that
\begin{equation}\label{pro6}
\sum_{k=0}^n \om_k^2=\al^2\sum_{k=0}^n \mu_k^2+2\al\be
\sum_{k=0}^n \mu_k\la_k+\be^2\sum_{k=0}^n \la_k^2
=\al^2+\be^2=|\ga|^2.
\end{equation}
By (\ref{MU_NU_LA}), $\mu_k,\la_k=O(n^{-1/4})$, hence
\begin{equation}\label{pro6a}
\omega_k^2=O\left(n^{-1/2}|\gamma|^2\right).
\end{equation}
 Partition all $k$'s
into $T$ groups $M_j$ so that for each group,
\begin{equation}\label{pro7}
\sum_{k\in M_j} \om_k^2\ge \frac{1}{2T}|\ga|^2.
\end{equation}
Then from (\ref{pro5}),
\begin{equation}\label{pro8}
|\Phi_n(\ga)|\le \prod_{j=1}^T\prod_{k\in M_j} \frac{1}{(1+a \om_k^2)^q}
\le \frac{1}{(1+a_0 \om_k^2)^{Tq}},\quad a_0=\frac{a}{2T}\,.
\end{equation}
Take $T=L/q$, then (\ref{pro1}) follows. Lemma \ref{univ1} is proved. 
\end{proof}

\begin{lemma}\label{univ2}
If $\f(s)$ satisfies (\ref{C1}), (\ref{C2}), then for any $L>0$
there exist $a_0,C>0$ and $N_0>0$ such that for all $n\ge N_0$,
\begin{equation}\label{pro9}
\left|\frac{\partial^j\Phi_n(\ga)}{\partial \be^j}\right|
\le\frac{C}{(1+a_0 |\ga|^2)^L}\,,\quad j=1,2,3.
\end{equation}
\end{lemma}

\begin{proof} Consider first $j=1$. From (\ref{pro2}),
\begin{equation}\label{pro21}
\left|\frac{\partial\Phi_n(\ga)}{\partial \be}\right|
=\left|\sum_{k=0}^n\la_k \f'(\om_k)
\prod_{l\not= k} \phi(\om_k)\right|\,.
\end{equation}
By repeating the proof of Proposition \ref{univ1} we obtain that
\begin{equation}\label{pro22}
\left|\prod_{l\not= k} \phi(\om_k)\right|\le \frac{1}{(1+a_0 |\ga|^2)^L}
\end{equation}
By (\ref{C3}), 
\begin{equation}\label{pro23}
|\f'(\om_k)|\le c_2|\om_k|,
\end{equation}
hence
\begin{equation}\label{pro24}
\sum_{k=0}^n |\la_k\f'(\om_k)|\le c_2\left(\sum_{k=0}^n\la_k^2\right)^{1/2}
\left(\sum_{k=0}^n\om_k^2\right)^{1/2}=c_2|\ga|.
\end{equation}
Thus,
\begin{equation}\label{pro25}
\left|\frac{\partial\Phi_n(\ga)}{\partial \be}\right|
\le\frac{c_2|\ga|}{(1+a_0 |\ga|^2)^L},
\end{equation}
which implies (\ref{pro9}) for $j=1$.

Consider $j=2$. From (\ref{pro2}),
\begin{equation}\label{pro26}
\frac{\partial^2\Phi_n(\ga)}{\partial \be^2}
=\sum_{k=0}^n\sum_{i\not=k}\la_i\la_k \f'(\om_i)\f'(\om_k)
\prod_{l\not= i,k} \phi(\om_k)+\sum_{k=0}^n\la_k^2 \f''(\om_k)
\prod_{l\not= k} \phi(\om_k)\,.
\end{equation}
By repeating the proof of Proposition \ref{univ1} we obtain that
\begin{equation}\label{pro27}
\left|\prod_{l\not= i,k} \phi(\om_k)\right|\le \frac{1}{(1+a_0 |\ga|^2)^L}\,.
\end{equation}
In addition, from (\ref{pro24}) we obtain that
\begin{equation}\label{pro28}
\left|\sum_{k=0}^n\sum_{i\not=k}\la_i\la_k \f'(\om_i)\f'(\om_k)\right|\le
\left(\sum_{k=0}^n|\la_k \f'(\om_k)|\right)^2\le c_2^2|\ga|^2.
\end{equation}
Therefore,
\begin{equation}\label{pro29}
\left|\sum_{k=0}^n\sum_{i\not=k}\la_i\la_k \f'(\om_i)\f'(\om_k)
\prod_{l\not= i,k} \phi(\om_k)\right|\le
\frac{c_2^2|\ga|^2}{(1+a_0 |\ga|^2)^L}\,.
\end{equation}
By (\ref{C2}) and (\ref{pro22}),
\begin{equation}\label{pro210}
\left|\sum_{k=0}^n\la_k^2 \f''(\om_k)
\prod_{l\not= k} \phi(\om_k)\right|
\le \frac{c_2}{(1+a_0 |\ga|^2)^L}\sum_{k=0}^n\la_k^2 
=\frac{c_2}{(1+a_0 |\ga|^2)^L}\,.
\end{equation}
Equation (\ref{pro26}) and estimates (\ref{pro29}), (\ref{pro210})
imply (\ref{pro9}) for $j=2$. The case $j=3$ is considered in the
same way. Lemma \ref{univ2} is proved.
\end{proof}

\begin{lemma}\label{univ3}
If $\f(s)$ satisfies (\ref{C1}), (\ref{C2}), then 
there exist $C>0$ and $N_0>0$ such that for all $n\ge N_0$,
\begin{equation}\label{univ3:1}
|\wh D_n(0,\eta;x)|\le \frac{C}{(1+|\eta|)^3}.
\end{equation}
\end{lemma}

{\it Proof.} Observe that by (\ref{char1}),
\begin{equation}\label{univ3:2}
\eta^k \wh D_n(0,\eta;x)=
\frac{(-i)^k}{(2\pi)^2}
\int_{\R^2}e^{-i\be\eta}\frac{\partial^k\Phi_n(\ga)}{\partial \be^k}
\,d\ga\,, 
\end{equation}
hence 
\begin{equation}\label{univ3:3}
(1+ |\eta|^3) |\wh D_n(0,\eta;x)|\le
\int_{\R^2}|\Phi_n(\ga)|\,d\ga+\int_{\R^2}
\left|\frac{\partial^3\Phi_n(\ga)}{\partial \be^3}\right|\,d\ga\,.
\end{equation}
From (\ref{pro4}) and (\ref{pro9}) we obtain that 
\begin{equation}\label{univ3:5}
(1+ |\eta|^3) |\wh D_n(0,\eta;x)|\le
\int_{\R^2}\frac{C}{(1+\ep_0 |\ga|^2)^L}\,d\ga\le C_0\,,
\end{equation}
if $L$ is taken greater than $1$. This implies (\ref{univ3:1}).
Lemma \ref{univ3} is proved.

Let $\kappa>0$ be a fixed small number,
\begin{equation}\label{dens0}
\kappa<\frac{1}{12}\,.
\end{equation}
 Set 
\begin{equation}\label{dens1}
\La_n=\{\ga:|\ga|\le n^\kappa\}.
\end{equation}

\begin{lemma}\label{univ4}
If $\f(s)$ satisfies (\ref{C1}), (\ref{C2}), then for any $L>0$
there exist $a_0>0$, $C>0$ and $N_0>0$ such that for all $n\ge N_0$,
\begin{equation}\label{dens2}
\sup_{\ga\in\La_n}\left|\Phi_n(\ga)-e^{-\frac{1}{2}|\ga|^2}
\right|\le \frac{C n^{-(1/4)+\kappa_0}}{(1+a_0|\ga|^2)^L},\quad
\kappa_0=3\kappa.
\end{equation}
\end{lemma}

{\it Proof.} Observe that
\begin{equation}\label{dens3}
\sum_{k=0}^n\om_k^2=|\ga|^2,
\end{equation}
and if $\ga\in\La_n$ then
\begin{equation}\label{dens3a}
\om_k=O(n^{-(1/4)+\kappa}).
\end{equation}
To prove (\ref{dens2}), let us write that
\begin{equation} \label{dens4}
\begin{aligned}
\Phi_n(\ga)-e^{-\frac{1}{2}\sum_{k=0}^n\om_k^2}&=
\prod_{k=0}^n \f(\om_k)-\prod_{k=0}^n e^{-\frac{1}{2}\om_k^2}\\
{}&=
\sum_{j=0}^n \left(\prod_{k=0}^{j-1} \f(\om_k)\right)
\left(\f(\om_j)-e^{-\frac{1}{2}\om_j^2}\right)
\prod_{k=j+1}^n e^{-\frac{1}{2}\om_k^2}\,.
\end{aligned}
\end{equation}
We have the estimate,
\begin{equation} \label{dens4a}
e^{-\frac{1}{2}x^2}\le \frac{1}{1+\frac{1}{2}x^2},
\end{equation}
hence similar to Lemma \ref{univ1} we obtain that
\begin{equation} \label{dens4b}
\left(\prod_{k=0}^{j-1} \f(\om_k)\right)
\prod_{k=j+1}^n e^{-\frac{1}{2}\om_k^2}\le \frac{1}{(1+a_0|\ga|^2)^L}.
\end{equation}
Due to (\ref{phi}), we have that as $s\to 0$,
$\f(s)=1-\frac{1}{2}s^2+O(|s|^3)$, hence there exists some constant
$C_0>0$ such that 
\begin{equation} \label{dens5}
\left|\f(\om_j)-e^{-\frac{1}{2}\om_j^2}\right|\le C_0|\om_j|^3\,,
\quad \ga\in\La_n\,.
\end{equation}
Thus, from (\ref{dens4}) we obtain that 
\begin{equation} \label{dens6}
\left|\Phi_n(\ga)-e^{-\frac{1}{2}\sum_{k=0}^n\om_k^2}\right|
\le \frac{C_0}{(1+a_0|\ga|^2)^L}\sum_{k=0}^n |\om_k|^3
\le \frac{C_0}{(1+a_0|\ga|^2)^L}(\sup_k |\om_k|)|\ga|^2\,.
\end{equation}
Hence,
\begin{equation} \label{dens7}
\left|\Phi_n(\ga)-e^{-\frac{1}{2}|\ga|^2}\right|
\le \frac{C n^{-(1/4)+\kappa_0}}{(1+a_0|\ga|^2)^L}\,.
\end{equation}
Lemma \ref{univ4} is proved.

\begin{lemma}\label{univ5}
If $\f(s)$ satisfies (\ref{C1}), (\ref{C2}), then 
there exist  $C>0$ and $N_0>0$ such that for all $n\ge N_0$,
\begin{equation}\label{dens8}
\sup_{\xi,\eta}\left|\wh D_n(\xi,\eta;x)-
\frac{1}{2\pi}\,e^{-\frac{1}{2}(\xi^2+\eta^2)}
\right|\le C n^{-(1/4)+\kappa_0},\quad
\kappa_0=3\kappa.
\end{equation}
\end{lemma}

{\it Proof.} From (\ref{pro1}), 
\begin{equation} \label{dens9}
\wh D_n(\xi,\eta;x)-
\frac{1}{2\pi}\,e^{-\frac{1}{2}(\xi^2+\eta^2)}
= \frac{1}{(2\pi)^2}\int_{-\infty}^\infty  \int_{-\infty}^\infty 
(\Phi(\ga)-e^{-\frac{1}{2}|\ga|^2})e^{-i\al\xi -i\be\eta}d\xi d\eta,
\end{equation}
hence (\ref{dens8}) follows from (\ref{dens2}). Lemma \ref{univ5} is proved.

{\it Proof of Theorem \ref{univer}.} By (\ref{RDENSITY}) we have that
\begin{equation} \label{dens10}
\frac{p_n(x)}{\sqrt n}-\frac{1}{\pi(1+x^2)}=
\frac{1}{1+x^2}\int_{-\infty}^\infty |\eta|\left(
\wh D_n(0,\eta;x)-\frac{1}{2\pi}\,e^{-\frac{1}{2}\eta^2}\right)d\eta.
\end{equation}
Let $\tau>0$ be an arbitrary number. By Lemma \ref{univ5},  
\begin{equation} \label{dens11}
\int_{-n^\tau}^{n^\tau} |\eta|\left|
\wh D_n(0,\eta;x)-\frac{1}{2\pi}\,e^{-\frac{1}{2}\eta^2}\right|d\eta
\le Cn^{2\tau-(1/4)+\kappa_0}.
\end{equation}
By Lemma \ref{univ3},
\begin{equation}\label{dens12}
\int_{|\eta|>n^\tau}|\eta||\wh D_n(0,\eta;x)|d\eta 
\le C\int_{|\eta|>n^\tau}\frac{|\eta| d\eta}{(1+|\eta|)^3}\le 2C
n^{-\tau} 
\end{equation}
Also, 
\begin{equation}\label{dens12b}
\int_{|\eta|>n^\tau}|\eta|e^{-\frac{1}{2}\eta^2}d\eta=
2e^{-\frac{1}{2}n^{2\tau}}. 
\end{equation}
Take $\tau=\frac{1}{12}-\frac{\kappa_0}{3}$. Then, combining the last three
estimates, we obtain that there exists $C>0$ such that
\begin{equation} \label{dens13}
\left|\frac{p_n(x)}{\sqrt n}-\frac{1}{\pi(1+x^2)}\right|\le
\frac{Cn^{-\frac{1}{12}+\frac{\kappa_0}{3}}}{1+x^2}=
\frac{Cn^{-\frac{1}{12}+\kappa}}{1+x^2},
\end{equation}
which implies (\ref{uni1}). 
Theorem \ref{univer} is proved.

\section{Scaling Near Zero}\label{scaling1}

In this section we will describe a crossover asymptotics
from (\ref{uni2}) to (\ref{uni1}).
The crossover takes place on a small scale of the order
of $n^{-1/2}$. Define the scaled variable $y$ as
\begin{equation} \label{z1}
y=n^{1/2}x.
\end{equation}
Consider in Proposition \ref{dens} the random variables 
\begin{equation}\label{z2}
g_n\left(\frac{y}{\sqrt n}\right)
=\sum_{k=0}^n\mu_k\left(\frac{y}{\sqrt n}\right)\,c_k\,,\qquad 
h_n\left(\frac{y}{\sqrt n}\right)
=\sum_{k=0}^n\la_k\left(\frac{y}{\sqrt n}\right)\,c_k\,.
\end{equation}
By (\ref{prop2}),
\begin{equation}\label{z3}
\begin{aligned}
\mu_k\left(\frac{y}{\sqrt n}\right)&= \frac{y^k}   
{n^{k/2}(1+\frac{y^2}{n})^{n/2}}\binom
  {n}{k}^{1/2}\,,
\quad
\la_k\left(\frac{y}{\sqrt n}\right)
= \mu_k\left(\frac{y}{\sqrt n}\right)
 (u-u_0)\frac{ n+y^2}{y}\,;\\
 u&=\frac{k}{n}\,,\quad  
u_0\equiv \frac{y^2}{n+y^2}\,,
\end{aligned}
\end{equation}
which gives that
\begin{equation}\label{z3a}
\la_k\left(\frac{y}{\sqrt n}\right)
= \mu_k\left(\frac{y}{\sqrt n}\right)
\left(\frac{k-y^2}{y}+\frac{ky}{n}\right).
\end{equation}
In particular,
\begin{equation}\label{z3b}
\begin{aligned}
\mu_0\left(\frac{y}{\sqrt n}\right)
&=\left(1+\frac{y^2}{n}\right)^{-n/2},\qquad 
\mu_1\left(\frac{y}{\sqrt n}\right)
=y\left(1+\frac{y^2}{n}\right)^{-n/2};\\
\la_0\left(\frac{y}{\sqrt n}\right)
&=-y\left(1+\frac{y^2}{n}\right)^{-n/2},\qquad 
\la_1\left(\frac{y}{\sqrt n}\right)
=\left(1-y^2\frac{n-1}{n}\right)
\left(1+\frac{y^2}{n}\right)^{-n/2}.
\end{aligned}
\end{equation}
As $n\to\infty$, we have the limits,
\begin{equation}\label{z4}
\begin{aligned}
\lim_{n\to\infty}\mu_k\left(\frac{y}{\sqrt n}\right)&= 
\frac{1}{\sqrt{k!}}\,y^ke^{-\frac{y^2}{2}}\equiv m_k(y),\\
\lim_{n\to\infty}\la_k\left(\frac{y}{\sqrt n}\right)
&=\frac{(k-y^2)}{\sqrt{k!}}\,y^{k-1}e^{-\frac{y^2}{2}}
\equiv l_k(y)=m_k(y)\,\frac{k-y^2}{y}\,. 
\end{aligned}
\end{equation}
Moreover, we have the following estimate of the error term.

\begin{lem}\label{mu} There is $C>0$ such that 
if $|y|\le n^{1/8}$ and $k\le n^{1/4}$, then
\begin{equation}\label{z4a}
\left|\mu_k\left(\frac{y}{\sqrt n}\right)-m_k(y)\right|
\le
\left\{
\begin{aligned}
{}& C\,\frac{y^4}{n}\,|m_k(y)|,\quad k=0,1,\\        
{}& C\,\frac{k^2+y^4}{n}\,|m_k(y)|,\quad k\ge 2,
\end{aligned}
\right.
\end{equation}
and 
\begin{equation}\label{z4aa}
\left|\la_k\left(\frac{y}{\sqrt n}\right)-l_k(y)\right|
\le
\left\{
\begin{aligned}
{}& C\,\frac{|y|+|y|^5}{n}\,|m_k(y)|,\quad k=0,1,\\        
{}& C\,\frac{k^3+y^6}{n|y|}\,|m_k(y)|,\quad k\ge 2,
\end{aligned}
\right.
\end{equation}
In addition, for $|y|\le n^{1/8}$  and all $k$,
\begin{equation}\label{z4b}
\left|\mu_k\left(\frac{y}{\sqrt n}\right)\right|
\le C\frac{|y|^k}{\sqrt{k!}}e^{-y^2/2}\,,\qquad
\left|\la_k\left(\frac{y}{\sqrt n}\right)\right|
\le \frac{|y|^k}{\sqrt{k!}}e^{-y^2/2}
\left[k\left(\frac{1}{|y|}+\frac{|y|}{n}\right)+1\right].         
\end{equation}
\end{lem}

{\it Proof.} From (\ref{z3}), (\ref{z4}) we have that
\begin{equation}\label{mu1}
\frac{\mu_k\left(\frac{y}{\sqrt n}\right)}{m_k(y)}
=\left[\prod_{j=0}^{k-1}\left(1-\frac{j}{n}\right)\right]^{1/2}
\frac{e^{y^2/2}}{(1+\frac{y^2}{n})^{n/2}},
\end{equation}
hence
\begin{equation}\label{mu2}
\left|\ln \frac{\mu_k\left(\frac{y}{\sqrt n}\right)}{m_k(y)}\right|
=\left|\frac{1}{2}\left[\sum_{j=0}^{k-1}\ln \left(1-\frac{j}{n}\right)\right]
+\frac{y^2}{2}-\frac{n}{2}\ln \left(1+\frac{y^2}{n}\right)\right|\le
C_0\left(\frac{k(k-1)}{n}+\frac{y^4}{n}\right),
\end{equation}
which gives (\ref{z4a}). From (\ref{z3a}),
\begin{equation}\label{mu3}
\frac{\la_k\left(\frac{y}{\sqrt n}\right)}{l_k(y)}
=\frac{\mu_k\left(\frac{y}{\sqrt n}\right)}{m_k(y)}
\left(1+\frac{y^2}{n(k-y^2)}\right)
=\left[1+O\left(\frac{k(k-1)}{n}+\frac{y^4}{n}\right)\right]
\left(1+\frac{y^2}{n(k-y^2)}\right),
\end{equation}
which implies (\ref{z4aa}).
To prove (\ref{z4b}) observe that if $|y|\le n^{1/8}$ then
\begin{equation} \label{z4c}
\left(1+\frac{y^2}{n}\right)^{-n/2}\le Ce^{-y^2/2}.
\end{equation}
Indeed,
\begin{equation} \label{z4d}
\frac{n}{2}\ln\left(1+\frac{y^2}{n}\right)
=\frac{y^2}{2}+O\left(\frac{y^4}{n}\right),
\end{equation}
hence (\ref{z4c}) follows. From (\ref{z3}) and (\ref{z4c}) we obtain
(\ref{z4b}), QED.

Observe that
\begin{equation} \label{z5}
\sum_{k=0}^\infty m_k(y)^2=\sum_{k=0}^\infty l_k(y)^2=1,
\quad \sum_{k=0}^\infty m_k(y)l_k(y)=0.
\end{equation}
Consider the random variables,
\begin{equation} \label{z7}
\begin{aligned}
g(y)&=\sum_{k=0}^\infty m_k(y)c_k=e^{-\frac{y^2}{2}}
\sum_{k=0}^\infty \frac{1}{\sqrt{k!}}\,y^k c_k,\\
h(y)&=\sum_{k=0}^\infty l_k(y)c_k=e^{-\frac{y^2}{2}}
\sum_{k=0}^\infty \frac{(k-y^2)}{\sqrt{k!}}\,y^{k-1} c_k.
\end{aligned}
\end{equation}
Let $D(\xi,\eta;y)$ be the joint distribution density
of $g(y)$ and $h(y)$. By the Kac-Rice formula the
density $\hat p(y)$
of the distribution of zeros of $g(y)$ is equal to
\begin{equation}\label{cross2}
\hat p(y)=\int_{-\infty}^\infty |\eta| D(0,\eta;y)d\eta.
\end{equation}
We will prove the following result.

\begin{theo}\label{crossover}
Let $f_n(x)$ be a random polynomial of degree $n$, as defined in
(\ref{POLY}). Let $p_n(x)$  
be the distribution density function of real zeros of $f_n(x)$, and
let $\phi(s)$ be 
the characteristic function of $c_k$. Assume that  for some $a,A>0$,
\begin{equation}\label{cross0}
\begin{aligned}
|\f(s)|\le\frac{1}{(1+a|s|)^6}\,;
\qquad 
\left|\frac{d^j\f(s)}{ds^j}\right|\le\frac{A}{(1+a|s|)^6}\,,
\quad j=1,2,3;\quad s\in\R.
\end{aligned}
\end{equation}
Then 
\begin{equation}\label{cross1}
\lim_{n\to\infty}\frac{p_n\left(\frac{y}{\sqrt n}\right)
}{\sqrt{n}} =\hat p(y)\,,
\end{equation}
uniformly in $y$ in the interval $|y|\le n^{1/8}$. In addition,
\begin{equation}\label{cross2a}
\lim_{n\to\infty}\frac{p_n(x)
}{\sqrt{n}}=\frac{1}{\pi(1+x^2)},
\end{equation}
uniformly in $x$ in the set $\{ x:\,n^{-3/8}\le |x|\le 1\}$.
\end{theo}

As a corollary of Theorems \ref{univer} and \ref{crossover} we will
prove the following result. 

\begin{theo}\label{cor_cross} Under the assumptions of Theorem
  \ref{crossover},
the average number of zeros of random
  polynomial  
(\ref{POLY}) is asymptotically equal to $\sqrt{n}$, so that
\begin{equation}\label{cross3}
\lim_{n\to\infty}\frac{1}{\sqrt n}\int_{-\infty}^\infty
p_n(x)dx=1.
\end{equation}
\end{theo}

\section{Proof of Theorems \ref{crossover}, 
\ref{cor_cross}}\label{scaling2} 

By (\ref{RDENSITY}), 
\begin{equation} \label{cr1}
\frac{p_n\left(\frac{y}{\sqrt n}\right)
}{\sqrt{n}} =\frac{1}{1+\frac{y^2}{n}}
\int_{-\infty}^\infty |\eta| D_n(0,\eta;y)d\eta\,,
\end{equation}
where $D_n(\xi,\eta;y)$ is the joint distribution density
of $g_n\left(\frac{y}{\sqrt n}\right)$ and 
$h_n\left(\frac{y}{\sqrt n}\right)$. Therefore, Theorem
\ref{crossover} will be proven if we prove that 
\begin{equation} \label{cr2}
\lim_{n\to\infty}\int_{-\infty}^\infty |\eta| D_n(0,\eta;y)d\eta
=\int_{-\infty}^\infty |\eta| D(0,\eta;y)d\eta.
\end{equation} 
Let $\Phi_n(\ga;y)$, where $\ga=(\al,\be)$,
 be the joint
characteristic function of $g_n\left(\frac{y}{\sqrt n}\right)$ and 
$h_n\left(\frac{y}{\sqrt n}\right)$. Then, by (\ref{z2}), 
\begin{equation} \label{cr3}
\Phi_n(\ga;y)=\prod_{k=0}^n\f(\om_{kn}(y)),
\end{equation}
where $\f$ is the characteristic function of $c_k$ and 
\begin{equation} \label{cr4}
\om_{kn}(y)=\mu_k\left(\frac{y}{\sqrt n}\right)\al+
\la_k\left(\frac{y}{\sqrt n}\right)\be.
\end{equation}
Let $\Phi(\ga;y)$ be the joint
characteristic function of $g\left(\frac{y}{\sqrt n}\right)$ and 
$h\left(\frac{y}{\sqrt n}\right)$. Then, by (\ref{z7}), 
\begin{equation} \label{cr5}
\Phi(\ga;y)=\prod_{k=0}^\infty\f(w_k(y)),
\end{equation}
where
\begin{equation} \label{cr6}
w_k(y)=m_k(y)\al+
l_k(y)\be.
\end{equation}
Observe that
\begin{equation} \label{cr6a}
\sum_{k=0}^n\om_{kn}(y)^2=\sum_{k=0}^\infty w_k(y)^2=|\ga|^2.
\end{equation}

\begin{lem} \label{lem_cr1a}
There exists $C>0$ such that for all $y$ in the interval
$|y|\le n^{1/8}$,
\begin{equation} \label{cr11}
\sum_{k=0}^\infty |\om_{kn}(y)-w_k(y)|\le C\frac{|\ga|}{ n}\,, 
\end{equation}
where we set $\om_{kn}(y)=0$ for $k>n$.
\end{lem}

{\it Proof.} By Lemma \ref{mu}, if $k\le n^{1/4}$ then for some $C_0>0$,
\begin{equation} \label{cr11a}
|\om_{kn}(y)-w_k(y)|\le
\left\{
\begin{aligned}
{}& C_0 \frac{|y|+|y|^5}{n}|y|^ke^{-y^2/2}|\ga|,\quad k=0,1,\\
{}& C_0 \frac{k^3+y^6}{n\sqrt{k!}}
|y|^{k-1}e^{-y^2/2}|\ga|,\quad k\ge 2,
\end{aligned}
\right.
\end{equation}
and for $k>n^{1/4}$,
\begin{equation} \label{cr11b}
|\om_{kn}(y)-w_k(y)|\le 
\frac{|y|^k}{\sqrt{k!}}e^{-y^2/2}
\left[k\left(\frac{1}{|y|}+\frac{|y|}{n}\right)+1\right]\,|\ga|.
\end{equation}
By summing up these inequalities over
$k=0,1,\ldots$, we obtain (\ref{cr11}), QED.

\begin{lem}\label {tail1}
If $|y|\le n^{1/8}$ then for some $A_0>0$,
\begin{equation} \label{cr12}
\left|\frac{\partial^j\Phi_n(\ga;y)}{\partial\be^j}\right|
\le \frac{A_0}{(1+a|\ga|)^{6-j}},
\quad n=1,2,\ldots; \quad j=0,1,2,3.
\end{equation}
\end{lem}

{\it Proof.} Consider $j=0$. From (\ref{cr3}),
\begin{equation} \label{cr10}
\begin{aligned}
|\Phi_n(\ga;y)|\le \prod_{k=0}^n|\f(\om_{kn}(y))|
\le \prod_{k=0}^n\frac{1}{(1+a|\om_{kn}(y)|^2)^3}
&\le \frac{1}{\left(1+a\sum_{k=0}^n|\om_{kn}(y)|^2\right)^3}\\
&\le \frac{1}{(1+a|\ga|^2)^3}\,,
\end{aligned}
\end{equation}
which implies (\ref{cr12}) for $j=0$. 
Let now $j=1$. Since $\f(0)=0$,
we obtain from (\ref{cross0}) that 
\begin{equation}\label{cr12a}
\left|\frac{d\f(s)}{ds}\right|\le\frac{A|s|}{(1+a|s|)^6}\,.
\end{equation}
From (\ref{cr3}),
\begin{equation}\label{cr13}
\frac{\partial\Phi_n(\ga)}{\partial \be}
=\sum_{k=0}^n\la_k \f'(\om_{kn})
\prod_{l\not= k} \phi(\om_{kn})\,,
\end{equation}
hence there exists $C>0$ such that
\begin{equation}\label{cr14}
\begin{aligned}
\left|\frac{\partial\Phi_n(\ga)}{\partial \be}\right|
&\le C\sum_{k=0}^n|\la_k|\, |\om_{kn}|
\prod_{l=0}^n\frac{1}{(1+a|\om_{ln}(y)|^2)^3}
\le \frac{C}{(1+a|\ga|^2)^3}\sum_{k=0}^n|\la_k|\, |\om_{kn}|\\
&\le \frac{C}{(1+a|\ga|^2)^3}
\left(\sum_{k=0}^n|\la_k|^2\right)^{1/2}
\left(\sum_{k=0}^n|\om_{kn}|^2\right)^{1/2}
=\frac{C|\ga|}{(1+a|\ga|^2)^3} ,
\end{aligned}
\end{equation}
which implies (\ref{cr12}) for $j=1$. The cases
$j=2,3$ are dealt similarly, QED.

By the same argument we prove  similar estimates for $\Phi(\ga;y)$:

\begin{lem} \label{tail2}
If $|y|\le n^{1/8}$ then for some $A_0>0$,
\begin{equation} \label{cr17}
\left|\frac{\partial^j\Phi(\ga;y)}{\partial\be^j}\right|
\le \frac{A_0}{(1+a|\ga|)^{6-j}},
\quad n=1,2,\ldots; \quad j=0,1,2,3.
\end{equation}
\end{lem}

Now we estimate the difference of $\Phi_n(\ga;y)$ and $\Phi(\ga;y)$.
 
\begin{lem} \label{diff}
There exist $C>0$ such that for all $|y|\le n^{1/8}$,
\begin{equation} \label{cr20}
|\Phi_n(\ga;y)-\Phi(\ga;y)|\le \frac{C|\ga|}{n(1+a|\ga|)^6}\,.
\end{equation}
\end{lem}

{\it Proof.} Observe that
\begin{equation} \label{cr21}
\begin{aligned}
\Phi_n(\ga;y)-\Phi(\ga;y)&=\prod_{k=0}^\infty\f(\om_k(y))
-\prod_{k=0}^\infty\f(w_k(y))\\
&=\sum_{j=0}^\infty \left(\prod_{k=0}^{j-1}\f(\om_k(y))\right)
\left(\f(\om_j(y)-\f(w_j(y)\right)
\left(\prod_{k=j+1}^{\infty}\f(w_k(y))\right).
\end{aligned}
\end{equation}
Therefore, for some $C>0$,
\begin{equation} \label{cr22}
\left|\Phi_n(\ga;y)-\Phi(\ga;y)\right|
\le \frac{C}{(1+a|\ga|)^6}\sum_{j=0}^\infty
|\om_{jn}(y)-w_j(y)|.
\end{equation}
By Lemma \ref{lem_cr1a} this implies that
\begin{equation} \label{cr23}
\left|\Phi_n(\ga;y)-\Phi(\ga;y)\right|
\le \frac{C_0|\ga|}{n(1+a|\ga|)^6},
\end{equation}
QED.

We apply Lemmas \ref{tail1}, \ref{tail2} to estimate the tail of
$D_n(\xi,\eta;y)$ and $D(\xi,\eta;y)$.

\begin{lem} \label{tail3}
There exists $C>0$ such that for all $|y|\le n^{1/8}$,
\begin{equation} \label{cr24}
|D_n(0,\eta;y)|,\;|D(0,\eta;y)|\le \frac{C}{(1+|\eta|)^3}\,,
\quad \eta\in\R.
\end{equation}
\end{lem}

From Lemma \ref{diff} we obtain the estimate of the difference of
$D_n(\xi,\eta;y)|$ and $D(\xi,\eta;y)|$:

\begin{lem} \label{diff1}
There exists $C>0$ such that for all $|y|\le n^{1/8}$,
\begin{equation} \label{cr25}
\sup_{\eta\in\R}|D_n(0,\eta;y)-D(0,\eta;y)|\le \frac{C}{n}\,.
\end{equation}
\end{lem}

{\it Proof of Theorem \ref{crossover}}.
By (\ref{cr1}) and (\ref{cross2}), 
\begin{equation} \label{cr26}
\left|\left(1+\frac{y^2}{n}\right)
 \frac{p_n\left(\frac{y}{\sqrt n}\right)}{\sqrt n}-\hat
  p(y)\right|
\le \int_{-\infty}^\infty 
\left |D_n(0,\eta;y)-D(0,\eta;y)\right|dy\,.
\end{equation}
Set $R=n^{1/3}$. Then by Lemma \ref{diff1},
\begin{equation} \label{cr26a}
 \int_{|\eta|\le R}
\left |D_n(0,\eta;y)-D(0,\eta;y)\right|dy
\le \frac{2CR}{n} =2Cn^{-2/3}\,, 
\end{equation}
and by Lemma \ref{tail3},
\begin{equation} \label{cr27}
 \int_{|\eta|\ge R}
\left |D_n(0,\eta;y)-D(0,\eta;y)\right|dy
\le C  R^{-2}=Cn^{-2/3}\,. 
\end{equation}
Thus, if $|y|\le n^{1/8}$ then
\begin{equation} \label{cr26b}
\left|\left(1+\frac{y^2}{n}\right)
 \frac{p_n\left(\frac{y}{\sqrt n}\right)}{\sqrt n}-\hat
  p(y)\right|
\le 3Cn^{-2/3}\,.
\end{equation}
This proves (\ref{cross1}).

To prove (\ref{cross2a}) observe that if $n^{-3/8}\le |x|\le 1$ then
by (\ref{MU_NU_LA}) there exists $C>0$ such that
\begin{equation}\label{MU_NU_LA2}
\max_{0\le k\le
  n}|\mu_k(x)|\le\frac{C}{n^{1/16}}\,,
\qquad
\max_{0\le k\le
  n}|\la_k(x)|\le\frac{C}{n^{1/16}}\,, 
\end{equation}
hence 
\begin{equation}\label{MU_NU_LA3}
\lim_{n\to\infty}\max_{0\le k\le
  n}|\mu_k(x)|=
\lim_{n\to\infty}
\max_{0\le k\le
  n}|\la_k(x)|=0\,. 
\end{equation}
Therefore, in this case  the proof of Theorem 
\ref{univer} is applicable and (\ref{cross2a})
follows. 

{\it Proof of Theorem \ref{cor_cross}}. By comparing (\ref{cross1})
with (\ref{cross2a}) we obtain that
\begin{equation}\label{cr27a}
\lim_{y\to\infty} \hat
p(y)=\frac{1}{\pi}\,,    
\end{equation}
hence by (\ref{cross1}), 
\begin{equation}\label{cr27b}
\begin{aligned}
\lim_{n\to\infty}n^{3/8}
\int_{|x|\le n^{-3/8}} p_n(x)dx&=\lim_{n\to\infty} n^{3/8}
\int_{|y|\le n^{1/8}}
p_n\left(\frac{y}{\sqrt n}\right)\frac{dy}{\sqrt n}\\&=
\lim_{n\to\infty}\frac{1}{n^{1/8}} \int_{|y|\le n^{1/8}} \hat
p(y)dy=\frac{1}{\pi}\,.
\end{aligned}    
\end{equation}
In addition, from (\ref{cross2a}) we obtain that
\begin{equation}\label{cr28}
\lim_{n\to\infty}
\frac{1}{\sqrt n}\int_{ n^{-3/8}\le |x|\le 1} p_n(x)dx=\int_{|x|\le 1}
\frac{1}{\pi(1+x^2)}dx=\frac{1}{2}\,.
\end{equation}
By combining these two relations we obtain that 
\begin{equation}\label{cr29}
\lim_{n\to\infty}
\frac{1}{\sqrt n}\int_{  |x|\le 1} p_n(x)dx=\frac{1}{2}\,.
\end{equation}
Observe that the probability distribution of zeros, $p_n(x)dx$, is
invariant with respect to the transformation
$x\to x^{-1}$. Indeed, the probability distribution of the
polynomial 
\begin{equation}\label{cr30}
x^nf_n(x^{-1})=\sum_{k=0}^n \sqrt{\binom{n}{k}}\,c_{n-k} x^k,
\end{equation}
coincides with the one of $f_n(x)$, because $c_k$'s are identically
distributed. Hence the distribution of zeros of $x^nf_n(x^{-1})$
coincides with the one of $f_n(x)$, so that it is invariant with
respect to the transformation $x\to x^{-1}$. Thus,
\begin{equation}\label{cr31}
\lim_{n\to\infty}
\frac{1}{\sqrt n}\int_{  |x|\ge 1} p_n(x)dx=
\lim_{n\to\infty}
\frac{1}{\sqrt n}\int_{  |x|\le 1} p_n(x)dx=\frac{1}{2}\,,
\end{equation}
and (\ref{cross3}) follows, QED.

\section{Existence and Universality of Limiting Correlation Functions}
\label{uni-corr1}

Let $f_n(x)$ be a random polynomial of degree $n$, as defined in
(\ref{POLY}), 
and let  ($x_1,\ldots,x_m$) be $m$ distinct points. We will assume that 
all $x_i\not=0$. To evaluate the $m$-point correlation function of
zeros we will 
use the following extension of the Kac-Rice formula (see \cite{BD},
\cite{BSZ2}): 
\begin{equation}\label{corr1}
K_{nm}(x_1,\ldots,x_m)
=\int_{-\infty}^\infty \cdots \int_{-\infty}^\infty
|\eta_1\cdots\eta_m|D_{nm}(0,\eta;x_1,\ldots,x_m)d\eta_1\cdots d\eta_m,
\end{equation}
where $D_{nm}(\xi,\eta;x_1,\cdots,x_m)$ is the joint 
distribution density function of the random vectors
$\xi=(f_n(x_1),\ldots,f_n(x_m))$ and $\eta=(f'_n(x_1),\ldots,f'_n(x_m))$.
By a change of variables, formula (\ref{corr1}) is first reduced to
\begin{equation}\label{corr2}
K_{nm}(x_1,\ldots,x_m)
=\prod_{i=1}^m\left[\frac{\z_n(x_i)}{\sg_n(x_i)}\right]
\int_{-\infty}^\infty \cdots \int_{-\infty}^\infty
|\eta_1\cdots\eta_m|\wt D_{nm}(0,\eta;x_1,\ldots,x_m)d\eta_1\cdots d\eta_m,
\end{equation}
where $\wt D_{nm}(0,\eta;x_1,\ldots,x_m)$ is the joint 
distribution density function of the random vectors
$$
\xi=(g_n(x_1),\ldots,g_n(x_m)),\quad \eta=(\wt g_n(x_1),\ldots,\wt g_n(x_m))
$$
[cf. (\ref{MDENSITY})], and then it is reduced to
\begin{equation}\label{corr3}
K_{nm}(x_1,\ldots,x_m)
=\prod_{i=1}^m\left[\frac{\sqrt n}{(1+x_i^2)}\right]
\int_{-\infty}^\infty \cdots \int_{-\infty}^\infty
|\eta_1\cdots\eta_m|\wh D_{nm}(0,\eta;x_1,\ldots,x_m)d\eta_1\cdots d\eta_m,
\end{equation}
where $\wh D_{nm}(\xi,\eta;x_1,\ldots,x_m)$ is the joint 
distribution density function of the random vectors
$$
\xi=(g_n(x_1),\ldots,g_n(x_m)),\quad \eta=(h_n(x_1),\ldots,h_n(x_m))
$$
[cf. (\ref{RDENSITY})]. We find now that
\begin{equation}\label{corr4}
\E g_n(x_i)=\E\left[\sum_{k=0}^n \mu_k(x_i)\,c_k\right]=0\,,
\quad
\E h_n(x_i)=\E\left[\sum_{k=0}^n \la_k(x_i)\,c_k\right]=0\,,
\end{equation}
and from (\ref{muk}), that
\begin{equation}\label{corr5}
\E g_n(x_i)g_n(x_j)=\sum_{k=0}^n \mu_k(x_i)\,\mu_k(x_j)
 = \left[\frac{1+x_i x_j}{(1+x_i^2)^{1/2}(1+x_j^2)^{1/2}}\right]^n 
\equiv \al_n(x_i,x_j)\,.
\end{equation}
Next, by (\ref{LA}) and (\ref{numu}),
\begin{equation}\label{corr6}
\la_k(x)=\frac{\nu_k(x)-(\nu(x),\mu(x))\mu_k(x)}{\sg(x)}\,;
\quad (\nu(x),\mu(x))=\frac{\sqrt n\,x}{\sqrt {1+nx^2}}\,,
\quad \sg(x)=\frac{1}{\sqrt {1+nx^2}}\,,
\end{equation}
and by (\ref{muk}),
\begin{equation}\label{corr7}
\begin{aligned}
\sum_{k=0}^n \mu_k(x_i)\,\nu_k(x_j)
&= \frac{\sqrt{n}\,(1+x_j^2)\,x_i}{\sqrt{1+nx_j^2}\,(1+x_ix_j)}
\,\al_n(x_i,x_j)\,,\\
\sum_{k=0}^n \nu_k(x_i)\,\nu_k(x_j)
&= \frac{(1+nx_ix_j)\,(1+x_i^2)\,(1+x_j^2)}
{\sqrt{1+nx_i^2}\,\sqrt{1+nx_j^2}\,(1+x_ix_j)^2}
\,\al_n(x_i,x_j)\,.
\end{aligned}
\end{equation}
This gives that 
\begin{equation}\label{corr8}
\begin{aligned}
\E g_n(x_i)h_n(x_j)&=\sum_{k=0}^n \mu_k(x_i)\,\la_k(x_j)
 = \sqrt{n}\,\frac{x_i-x_j}{1+x_ix_j}\,\al_n(x_i,x_j)\,,\\
\E h_n(x_i)h_n(x_j)&=\sum_{k=0}^n \la_k(x_i)\,\la_k(x_j)
 = \frac{(1+x_i^2)\,(1+x_j^2)-n(x_i-x_j)^2}
{(1+x_ix_j)^2}\,\al_n(x_i,x_j)\,.
\end{aligned}
\end{equation}
Let us make the change of variable,
\begin{equation}\label{corr9}
\theta=\arctan x.
\end{equation}
Then formulae (\ref{corr5}),
(\ref{corr8}) simplify to
\begin{equation}\label{corr10}
\begin{aligned}
\E g_n(\theta_i)g_n(\theta_j)&=\cos^n(\theta_i-\theta_j)\,,\\
\E g_n(\theta_i)h_n(\theta_j)&=\sqrt{n}\,\tan(\theta_i-\theta_j)\,
\cos^n(\theta_i-\theta_j)\,,\\
\E h_n(\theta_i)h_n(\theta_j)&=\left[\frac{1}{\cos^2(\theta_i-\theta_j)}-
n\,\tan^2(\theta_i-\theta_j)\right]
\cos^n(\theta_i-\theta_j)\,.
\end{aligned}
\end{equation}
To get a proper scaling we fix a  $\theta^0$, the reference point,
 and set
\begin{equation}\label{corr10a}
\theta=\theta^0+\frac{y}{\sqrt n}\,,
\end{equation}
where $y$ is a scaled variable. Then (\ref{corr10}) reduces to
\begin{equation}\label{corr10b}
\begin{aligned}
\E g_n(\theta^0+\frac{y_i}{\sqrt n})
g_n(\theta^0+\frac{y_j}{\sqrt n})&=\cos^n\frac{y_i-y_j}{\sqrt n}
\equiv a_n(y_i,y_j)\,,\\
\E g_n(\theta^0+\frac{y_i}{\sqrt n})
h_n(\theta^0+\frac{y_j}{\sqrt n})&=\sqrt{n}\,\tan\frac{y_i-y_j}{\sqrt n}\,
\cos^n\frac{y_i-y_j}{\sqrt n}\equiv b_n(y_i,y_j)\,,\\
\E h_n(\theta^0+\frac{y_i}{\sqrt n})
h_n(\theta^0+\frac{y_j}{\sqrt n})
&=\left[\frac{1}{\cos^2\frac{y_i-y_j}{\sqrt n}}-
n\,\tan^2\frac{y_i-y_j}{\sqrt n}\right] 
\cos^n\frac{y_i-y_j}{\sqrt n}\equiv c_n(y_i,y_j)\,.
\end{aligned}
\end{equation}
As $n\to\infty$,
\begin{equation}\label{corr11}
\begin{aligned}
\lim_{n\to\infty} a_n(y_i,y_j)&=e^{-(y_i-y_j)^2/2}\equiv
a(y_i,y_j)\,,\\ 
\lim_{n\to\infty} b_n(y_i,y_j)&=(y_i-y_j)e^{-(y_i-y_j)^2/2}\equiv
b(y_i,y_j)\,,\\ 
\lim_{n\to\infty}
c_n(y_i,y_j)&=[1-(y_i-y_j)^2]e^{-(y_i-y_j)^2/2}\equiv c(y_i,y_j)\,,\\ 
\end{aligned}
\end{equation}
More precisely, it follows from (\ref{corr10}) that as $n\to\infty$,
\begin{equation}\label{corr11a}
\begin{aligned}
a_n(y_i,y_j)&=a(y_i,y_j)+O(n^{-1})\,,\quad
b_n(y_i,y_j)=b(y_i,y_j)+O(n^{-1})\,,\\
c_n(y_i,y_j)&=c(y_i,y_j)+O(n^{-1})\,.
\end{aligned}
\end{equation}
From (\ref{prop2}) we obtain that 
\begin{equation}\label{corr11b}
\begin{aligned}
\mu_k(\tan\theta)&= \frac{\tan^k\theta}
{(1+\tan^2\theta)^{n/2}}{\binom 
{n}{k}}^{1/2}=\sin^k\theta\cos^{n-k}\theta{\binom
{n}{k}}^{1/2}\,,\\ 
\la_k(\tan\theta)&=\mu_k(\tan\theta)\frac{\sqrt n
  (u-u_0)}{\sin\theta\cos\theta}\,, 
\end{aligned}
\end{equation}
hence formula (\ref{corr3}) reduces, under the change of variable
(\ref{corr9}), to the following.

\begin{prop}\label{corr_s}
The $m$-point correlation function of the scaled zeros,
$\tau_j\equiv\sqrt n(\arctan x_j-\theta^0)$,  is given by the formula 
\begin{equation}\label{corr12}
K_{nm}(y_1,\ldots,y_m)
=\int_{-\infty}^\infty \cdots \int_{-\infty}^\infty
|\eta_1\cdots\eta_m| D_{nm}(0,\eta;y_1,\ldots,y_m)d\eta_1\cdots d\eta_m,
\end{equation}
where $D_{nm}(\xi,\eta;y_1,\ldots,y_m)$ is the density of the joint
distribution of the random vectors 
$$
\xi=(g_n(y_1),\ldots,g_n(y_m)),\quad \eta=(h_n(y_1),\ldots,h_n(y_m))\,,
$$
where
\begin{equation}\label{corr13}
g_n(y)=\sum_{k=0}^n\mu_k(y)\,c_k\,,\quad
h_n(y)=\sum_{k=0}^n\la_k(y)\,c_k\,,
\end{equation}
and  
\begin{equation}\label{corr13a}
\begin{aligned}
\mu_k(y)&=\sin^k\left(\theta^0+\frac{y}{\sqrt n}\right)
\cos^{n-k}\left(\theta^0+\frac{y}{\sqrt n}\right){\binom{n}{k}}^{1/2},\\
\la_k(y)&=\frac{\mu_k(y)\sqrt n(u-u_0)}
{\sin\left(\theta^0+\frac{y}{\sqrt n}\right)
\cos\left(\theta^0+\frac{y}{\sqrt n}\right)};
\quad u=\frac{k}{n},\quad u_0=\sin^2\left(\theta^0+\frac{y}{\sqrt n}\right).
\end{aligned}
\end{equation}
\end{prop}

We can formulate now our main result. Consider a Gaussian
vector random field $(g(y),h(y))$ on the line such that
\begin{align}
&\E g(y)=\E h(y)=0; \nonumber\\
& \E g(y_i)g(y_j)=e^{-(y_i-y_j)^2/2}\equiv
a(y_i,y_j);\\
\label{corr15}
&\E g(y_i)h(y_j)=(y_i-y_j)e^{-(y_i-y_j)^2/2}\equiv
b(y_i,y_j)\,,\\
\label{corr16}
&\E h(y_i)h(y_j)=[1-(y_i-y_j)^2]e^{-(y_i-y_j)^2/2}\equiv c(y_i,y_j)\,.
\end{align}
It is realized as
\begin{equation}\label{gh1}
g(y)=e^{-y^2/2}\sum_{k=0}^\infty \frac{1}{\sqrt{k!}}\,c_k\,y^k\,,\quad
h(y)=g'(y)\,,
\end{equation}
where $c_k$ are independent standard Gaussian random variables.
Observe that the random series in
(\ref{gh1}) converges a.s., and it defines $g(y)$ as an entire function.
Let $D_m(\xi,\eta;y_1,\ldots,y_m)$ be the (Gaussian) joint 
 distribution density
of the vectors 
$$
\xi=(g(y_1),\ldots,g(y_m)),\quad
\eta=(h(y_1),\ldots,h(y_m)).
$$

\begin{theo}\label{univ_corr} Assume that $\f(s)$, the characteristic
  function of $c_k$, is an infinitely differentiable function such that
for some $a,q>0$,
\begin{equation}\label{pr2}
|\f(s)|\le\frac{1}{(1+a s^2)^q}\,.
\end{equation}
and for any $j\ge 2$ there exists
$c_j>0$ such that
\begin{equation}\label{pr2b}
\left|\frac{d^j\f(s)}{ds^j}\right|<c_j\,.
\end{equation}
Assume that the reference point $\theta^0\not=0$
 and $y_i\not=y_j$ for $i\not= j$.
Then for every $m\ge 1$, there exists the limit, 
\begin{equation}\label{corr17}
\lim_{n\to\infty} K_{nm}(y_1,\ldots,y_m)=
 K_m(y_1,\ldots,y_m)\,,
\end{equation}
where
\begin{equation}\label{corr18}
K_m(y_1,\ldots,y_m)
=\int_{-\infty}^\infty \cdots \int_{-\infty}^\infty
|\eta_1\cdots\eta_m| D_m(0,\eta;y_1,\ldots,y_m)d\eta_1\cdots d\eta_m,
\end{equation} 
\end{theo}

\begin{rems}
1. We will prove, in fact, that for any $\de>0$ and any $\ep>0$
there exists $C>0$ such that for any $\theta^0$ and 
$y=(y_1,\ldots,y_m)$ such
that $\de<|\theta^0|<\frac{\pi}{2}-\de$ and
$\de<|y_i-y_j|$, $\,i\not=j$,   
\begin{equation}\label{remark}
\left|K_{nm}(y_1,\ldots,y_m)- K_m(y_1,\ldots,y_m)\right|\le
  Cn^{-1/4+\ep}\,, 
\end{equation}
This estimates the rate of convergence of $K_{nm}(y_1,\ldots,y_m)$ to
$K_m(y_1,\ldots,y_m)$. 2. Conditions (\ref{pr2}), (\ref{pr2b})
are fulfilled for any density $r(t)$ of the class $\dcal_{\infty}$. 
\end{rems}

We will prove Theorem \ref{univ_corr} by showing  that 
$D_{nm}(\xi,\eta;y)$ converges to $D_m(\xi,\eta;y)$,
where $y=(y_1,\dots,y_m)$,
in an appropriate sense,
so that we will be able to prove that the integral in 
(\ref{corr12}) converges to the one in (\ref{corr18}).

\section{Proof of Theorem \ref{univ_corr} }\label{uni-corr2}

Let 
\begin{equation} \label{prove1}
\Phi_n(\ga) = \Phi_n(\ga;y) = \int_{-\infty}^\infty \cdots
\int_{-\infty}^\infty  
D_{nm}(\xi,\eta;y)e^{i(\al,\xi) + i(\be,\eta)}d\xi d\eta,
\quad y=(y_1,\ldots,y_m),
\end{equation}
be the characteristic function of the random vector $(\xi,\eta)$, where
\begin{equation}
\al=(\al_1,\ldots,\al_m), \quad \be=(\be_1,\ldots,\be_m),\quad 
\ga=(\al,\be)=(\al_1,\ldots,\al_m,\be_1,\ldots,\be_m)\,.
\end{equation}
From (\ref{corr13}) we have that
\begin{equation} \label{prove2}
\Phi_n(\ga)
= \prod_{k=0}^n \phi(\om_k)\,,
\end{equation}
where $\f$ is the characteristic function of $c_k$ and
\begin{equation}\label{pr1}
\om_k= \sum_{i=1}^m\left[\mu_k(y_i)\al_i+ \la_k(y_i)\be_i\right].
\end{equation}
We will prove the following basic lemma.

\begin{lemma}\label{basic}
If $\f(s)$ satisfies (\ref{pr2}), then for any $L>0$
there exist $a_0>0$ and $N_0>0$ such that for all $n\ge N_0$,
\begin{equation}\label{pr3}
|\Phi_n(\ga)|\le\frac{1}{(1+a_0 |\ga|^2)^L}\,.
\end{equation}
\end{lemma}

A proof of Lemma \ref{basic} is given in the Appendix below.
We will prove the following addition to the basic lemma.

\begin{lemma}\label{basic_add}
If $\f(s)$ satisfies (\ref{pr2}) and (\ref{pr2b})
then for any $L>0$ 
there exist $a_0>0$ and $N_0>0$ such that for 
any multi-index $k=(k_1,\ldots,k_{2m})$ there exists
$C_k>0$ such that for all $n\ge N_0$,
\begin{equation}\label{pr3a}
|D^k\Phi_n(\ga)|\le\frac{C_k}{(1+a_0 |\ga|^2)^L}\,,
\qquad D^k\equiv \frac{\partial^{k_1+\ldots+k_{2m}}}{\partial
    \gamma_1^{k_1} \ldots \partial \gamma_{2m}^{k_{2m}}}\,.
\end{equation}
\end{lemma}

A proof of Lemma \ref{basic_add} is given in the Appendix below.
Lemma \ref{basic_add} implies the following useful estimate.

\begin{lemma}\label{D}
If $\f(s)$ satisfies (\ref{pr2}) and (\ref{pr2b})
then for any $\,K>0$  
there exist $C>0$ and $N_0>0$ such that  for all  $n\ge N_0$,
\begin{equation}\label{D1}
|D_{nm}(\xi,\eta;y)|\le\frac{C}{(1+ |\xi|^2+|\eta|^2)^K}\,.
\end{equation}
\end{lemma}

{\it Proof.} Observe that by (\ref{prove1}),
\begin{equation}\label{D2}
(1+ |\xi|^2+|\eta|^2)^K D_{nm}(\xi,\eta;y)=
\frac{1}{(2\pi)^{2m}}
\int_{\R^{2m}}e^{-i(\al,\xi)-i(\be,\eta)}(1-\De)^K\Phi_n(\ga)\,d\ga\,, 
\end{equation}
where $\De$ is the Laplacian, hence 
\begin{equation}\label{D3}
(1+ |\xi|^2+|\eta|^2)^K |D_{nm}(\xi,\eta;y)|\le
\int_{\R^{2m}}|(1-\De)^K\Phi_n(\ga)|\,d\ga\,.
\end{equation}
From (\ref{pr3a}) we obtain that there exists a constant $C_0$ such
that
\begin{equation}\label{D4}
|(1-\De)^K\Phi_n(\ga)|\le \frac{C}{(1+a_0 |\ga|^2)^L}\,,
\end{equation}
hence
\begin{equation}\label{D5}
(1+ |\xi|^2+|\eta|^2)^K |D_{nm}(\xi,\eta;y)|\le
\int_{\R^{2m}}\frac{C}{(1+a_0 |\ga|^2)^L}\,d\ga\le C\,,
\end{equation}
if $L$ is taken greater than $m$. This implies (\ref{D1}).
Lemma \ref{D} is proved.

\begin{rem} \label{uniform}
The constants $a_0$ in (\ref{pr3}) and (\ref{pr3a}), $C_k$ in
  (\ref{pr3a}), and $C$ in (\ref{D1}) are uniform with respect to 
$\theta^0$ and $y$, assuming that $\de<|\theta^0|<\frac{\pi}{2}-\de$
and $|y_i-y_j|>\de$,  $i\not=j$, for some $\de>0$.
\end{rem}

{\it Proof of Theorem \ref{univ_corr}}. Let $\kappa>0$ be a small
fixed number, 
\begin{equation}\label{kappa}
\kappa<\frac{1}{8(2m+3)}\,.
\end{equation}
 Set
\begin{equation}\label{La_n1}
\La_n=\{\ga \big ||\ga| \le n^{\kappa}\}\,.
\end{equation}
By (\ref{MU_NU_LA}), there exists a constant $C_0>0$ such that
\begin{equation} \label{prove3}
\max_{k}|\mu_k(y_i)| \le C_0 n^{-1/4}, \quad  \max_{k}|\la_k(y_i)| \le
C_0 n^{-1/4}, 
\quad i = 1,\ldots,m \,.
\end{equation}
\noindent
Therefore, for each $\ga \in \Lambda_n$, we have that 
\begin{equation} \label{prove4}
|\omega_k| \le \sum_{i=1}^m(|\mu_k(y_i)\al_i| + |\la_k(y_i)\be_i|) \le
 C_1 n^{-1/4+\kappa},\quad C_1=2mC_0\,, 
\end{equation}
hence
\begin{equation}\label{prove5}
\sum_{k=0}^n |\om_k^3| \le C_1n^{-1/4+\kappa}\sum_{k=0}^n \om_k^2.
\end{equation}
From (\ref{pr1}) we have that
\begin{equation}\label{prove6}
\sum_{k=0}^n \omega_k^2 = \ga\De_n\ga^{T}\,,
\end{equation}
where $\ga^T$ is the transpose vector of $\ga$, and
\begin{equation} \label{MAT}
\De_n=\begin{pmatrix}
A_n & B_n \\
B_n^T & C_n
\end{pmatrix},
\end{equation}
with
\begin{equation}
A_n=\left(a_n(y_i,y_j)\right)_{i,j=1}^m,
\quad B_n=\left(b_n(y_i,y_j)\right)_{i,j=1}^m,
\quad C_n=\left(c_n(y_i,y_j)\right)_{i,j=1}^m.
\end{equation}
By (\ref{corr11a}), as $n\to\infty$,
\begin{equation}\label{prove7}
\De_n=\De+O(n^{-1})\,,
\end{equation}
where
\begin{equation} \label{MAT2}
\De=\begin{pmatrix}
A & B \\
B^T & C
\end{pmatrix},
\end{equation}
with
\begin{equation}
A=\left(a(y_i,y_j)\right)_{i,j=1}^m,
\quad B=\left(b(y_i,y_j)\right)_{i,j=1}^m,
\quad C=\left(c(y_i,y_j)\right)_{i,j=1}^m.
\end{equation}
The matrix $\De$ is invertible \cite{BD},
cf. (\ref{gh1}) above.
From (\ref{prove6}) and (\ref{prove7}) we have that if $\ga\in\La_n$
then 
\begin{equation} \label{prove8a}
\sum_{k=0}^n \omega_k^2 = \ga\De\ga^{T}+O(n^{-1+2\kappa})\,.
\end{equation}
We will prove the following lemma.

\begin{lemma}\label{La_n}
There exist $C>0$ and $N_0>0$ such that for all $n>N_0$,
\begin{equation} \label{Lan1}
\sup_{\ga\in\La_n}\left|\Phi_n(\ga)
-e^{-\frac{1}{2}\ga\De\ga^{T}}\right| 
\le Cn^{-1/4+\kappa_0}\,,\quad
\kappa_0=3\kappa\,.
\end{equation}
\end{lemma}

{\it Proof.} We will prove first that
\begin{equation} \label{Lan2}
\sup_{\ga\in\La_n}\left|\Phi_n(\ga)
-e^{-\frac{1}{2}\sum_{k=0}^n\om_k^2}\right|
\le Cn^{-1/4+\kappa_0}\,.
\end{equation}
Then we will use (\ref{prove8a}). To prove (\ref{Lan2}), let us write
that 
\begin{equation} \label{Lan3}
\begin{aligned}
\Phi_n(\ga)-e^{-\frac{1}{2}\sum_{k=0}^n\om_k^2}&=
\prod_{k=0}^n \f(\om_k)-\prod_{k=0}^n e^{-\frac{1}{2}\om_k^2}\\
{}&=
\sum_{j=0}^n \left(\prod_{k=0}^{j-1} \f(\om_k)\right)
\left(\f(\om_j)-e^{-\frac{1}{2}\om_j^2}\right)
\prod_{k=j+1}^n e^{-\frac{1}{2}\om_k^2}\,.
\end{aligned}
\end{equation}
Due to (\ref{CK}), we have that as $s\to 0$,
$\f(s)=1-\frac{1}{2}s^2+O(|s|^3)$, hence there exists some constant
$C_0>0$ such that 
\begin{equation} \label{Lan4}
\left|\f(\om_j)-e^{-\frac{1}{2}\om_j^2}\right|\le C_0|\om_j|^3\,,
\quad \ga\in\La_n\,.
\end{equation}
In addition, $|\f(s)|\le 1$, hence from (\ref{Lan3}) we obtain that 
\begin{equation} \label{Lan5}
\left|\Phi_n(\ga)-e^{-\frac{1}{2}\sum_{k=0}^n\om_k^2}\right|
\le C_0\sum_{k=0}^n |\om_k|^3\,.
\end{equation}
From (\ref{prove5}) and (\ref{prove6}) we obtain now that there exists some constant $C_1$ 
such that
\begin{equation} \label{Lan6}
\left|\Phi_n(\ga)-e^{-\frac{1}{2}\sum_{k=0}^n\om_k^2}\right|
\le C_1 n^{-1/4+\kappa_0}\,.
\end{equation}
Finally, from (\ref{prove8a}) we have that
\begin{equation} \label{Lan7}
\left|e^{-\frac{1}{2}\sum_{k=0}^n\om_k^2}-e^{-\frac{1}{2}\ga\De\ga^{T}}\right|
\le C_2 n^{-1+2\kappa}\,.
\end{equation}
Combining (\ref{Lan6}) with (\ref{Lan7}) we obtain (\ref{Lan1}).
Lemma \ref{La_n} is proved.

From Lemmas \ref{basic} and \ref{La_n} we will derive
the following lemma.  

\begin{lemma}\label{Dnm}
There exist $C>0$ and $N_0>0$ such that for all $n>N_0$, 
\begin{equation} \label{Dnm1}
\sup_{\xi,\eta}\left|D_{nm}(\xi,\eta;y)
-D_m(\xi,\eta;y)\right|\le Cn^{-1/4+\kappa_1}\,,
\quad \kappa_1=(2m+3)\kappa\,.
\end{equation}
\end{lemma}

{\it Proof.} From (\ref{prove1}) and the definition of $D_m$
we have that
\begin{equation} \label{Dnm2}
\begin{aligned}
D_{nm}(\xi,\eta;y)
&-D_m(\xi,\eta;y)
=\frac{1}{(2\pi)^{2m}}\int_{\R^{2m}}e^{-i(\xi,\al)-i(\eta,\be)}
\left(\Phi_n(\ga)-e^{-\frac{1}{2}\ga\De\ga^{T}}\right)\,d\ga\,.
\end{aligned}
\end{equation}
From Lemma \ref{La_n} we obtain that there exists $C_0>0$ such that
\begin{equation} \label{Dnm3}
\left|\int_{\La_n}e^{-i(\xi,\al)-i(\eta,\be)}
\left(\Phi_n(\ga)-e^{-\frac{1}{2}\ga\De\ga^{T}}\right)\,d\ga\right|
\le C_0 n^{-1/4+\kappa_0}\,\text{Vol}\,\La_n= C_1 n^{-1/4+\kappa_1}\,.
\end{equation}
By Lemma \ref{basic} there exists $C_0>0$ such that
\begin{equation} \label{Dnm4}
\left|\int_{\R^{2m}\setminus\La_n}e^{-i(\xi,\al)-i(\eta,\be)}
\Phi_n(\ga)\,d\ga\right|
\le C_0 \int_{\R^{2m}\setminus\La_n}\frac{1}{(1+a_0|\ga|^2)^L}\,d\ga
\le C_1 n^{-1},
\end{equation}
if we choose $L=L(\kappa)$ sufficiently large. Since
$e^{-\frac{1}{2}\ga\De\ga^{T}}$ 
is a nondegenerate Gaussian kernel, there exists, obviously, $C_0>0$
such that 
\begin{equation} \label{Dnm5}
\left|\int_{\R^{2m}\setminus\La_n}e^{-i(\xi,\al)-i(\eta,\be)}
e^{-\frac{1}{2}\ga\De\ga^{T}}\,d\ga\right|
\le \int_{\R^{2m}\setminus\La_n}
e^{-\frac{1}{2}\ga\De\ga^{T}}\,d\ga
\le C_0 n^{-1}.
\end{equation}
Combining estimates (\ref{Dnm3})-(\ref{Dnm5}) with equation (\ref{Dnm2})
we obtain (\ref{Dnm1}). Lemma \ref{Dnm} is proved.

We will prove Theorem \ref{univ_corr} from  Lemmas \ref{D}
and \ref{Dnm}. Let $\tau>0$ be a fixed small number.
By (\ref{corr12}) and (\ref{corr18}),
\begin{equation}\label{uc1}
\begin{aligned}
K_{nm}(y)&-K_m(y)
=\int_{\R^m}
|\eta_1\cdots\eta_m| \left[D_{nm}(0,\eta;y)-D_m(0,\eta;y)
\right]\,d\eta\,,\quad y=(y_1,\dots,y_m).
\end{aligned}
\end{equation} 
By Lemma \ref{Dnm},
\begin{equation}\label{uc2}
\begin{aligned}
\int_{\{\eta\,:\,|\eta|\le n^{\tau}\}}
&|\eta_1\cdots\eta_m|
\left|D_{nm}(0,\eta;y)-D_m(0,\eta;y) 
\right|\,d\eta \\
&\le C n^{-1/4+\kappa_1}\,\int_{\{\eta\,:\,|\eta|\le n^{\tau}\}}
|\eta_1\cdots\eta_m|\,d\eta\le C_1 n^{-1/4+\tau_1}\,,\quad
\tau_1=\kappa_1+2m\tau\,.
\end{aligned}
\end{equation} 
By Lemma \ref{D},
\begin{equation}\label{uc3}
\begin{aligned}
\int_{\{\eta\,:\,|\eta|\ge n^{\tau}\}}
&|\eta_1\cdots\eta_m D_{nm}(0,\eta;y)|\,d\eta\\ 
&\le C\,\int_{\{\eta\,:\,|\eta|\ge n^{\tau}\}}
|\eta_1\cdots\eta_m|\,\frac{1}{(1+|\eta|^2)^K}\,d\eta
\le C_1 n^{-1}\,,
\end{aligned}
\end{equation} 
if we take $K$ sufficiently large. Since $D_m$ is a Gaussian kernel,
there exists, obviously, $C_0>0$ such that
\begin{equation}\label{uc4}
\begin{aligned}
\int_{\{\eta\,:\,|\eta|\ge n^{\tau}\}}
&|\eta_1\cdots\eta_m D_m(0,\eta;y)|\,d\eta 
\le C_0 n^{-1}\,,
\end{aligned}
\end{equation}
From (\ref{uc1})-(\ref{uc4}) we obtain that there exists $C>0$ such that
\begin{equation}\label{uc5}
|K_{nm}(y)-K_m(y)|\le Cn^{-1/4+\tau_1}\,.
\end{equation} 
Since $\tau_1=(2m+3)\kappa+2m\tau$ can be made as small as we want, 
Theorem \ref{univ_corr} is proved.

\section{Conclusion}\label{conclusion} 

In the present work we have proved
that the correlation functions of real zeros of a random polynomial of 
form (\ref{POLY}) have a universal scaling limit 
if we stay away from the
origin. The method of the proof is based on the Kac-Rice type
 formula for the correlation functions and on the convergence of the 
Kac-Rice kernel to a
universal Gaussian limit. The convergence to the  Gaussian
kernel is established as a local central limit theorem for not
identically distributed multivariate random variables, with some
appropriate additional estimates. The method of the proof is rather
general and it can be extended to 
ensembles of multivariate random polynomials. Let us consider briefly
these extensions.

{\bf Real zeros of non-Gaussian multivariate random polynomials.} Let
$x=(x_0,\dots,x_d)\in\R^{d+1}$. Consider a random homogeneous
multivariate polynomial in $x$, 
\begin{equation}\label{con1}
f_n(x)=\sum_{|k|=n}\sqrt{\binom{n}{k}}\,c_kx^k, 
\end{equation} 
where 
\begin{equation}\label{con1a}
k=(k_0,\dots,k_d)\in \Z_+^{d+1},\quad
|k|=k_0+\dots+k_d,\quad   x^k=x_0^{k_0}\dots x_d^{k_d},
\quad \binom{n}{k}=\frac{n!}{k_0!\dots  k_d!}, 
\end{equation} 
$\Z_+=\{n\in \Z,\;n\ge 0\}$, 
and $c_k$ are identically distributed real random
variables. Assume that
\begin{equation}\label{con2}
\E c_k=0;\quad \E c_k^2=1.
\end{equation}
In the case when $\{c_k\}$ are independent
standard Gaussian random variables, we obtain the Gaussian SO$(d+1)$
ensemble. Consider $p$ independent copies, $f=(f_{n1}(x),\dots
f_{np}(x))$, 
of polynomial (\ref{con1}), $p\le d$, 
and the set of common zeros of these polynomials,
\begin{equation}\label{con3}
Z_f=\{x:\,f_{n1}(x)=\dots=f_{np}(x)=0\}.
\end{equation}
Since $f_{nj}(x)$ are homogeneous polynomials, we can view
$Z_f$ as a real algebraic variety in the projective space $\RP^d$. 
Assume that the distribution of $c_k$ is Lebesgue absolutely
continuous, with a smooth density.
Let $K_{nm}(x^{(1)},\dots,x^{(m)})$ be the $m$-point correlation
function of the common zeros, see \cite{BSZ1}-\cite{BSZ4}.
In the case when $\{c_k\}$ are independent
standard Gaussian random variables, the correlation functions are
SO$(d+1)$-invariant, and the average number of the common real zeros
is the square root of the total number of common complex zeros (see
\cite{EK}, \cite {Kos}, \cite {ShSm}).  
Our approach enables us to prove the following multivariate 
extension of Theorem \ref{univ_corr} above.    
Consider the random Gaussian multivariate analytic function of $d$
variables, 
\begin{equation}\label{con4}
g(y)=e^{-|y|^2/2}\sum_{k\in\Z^d_+} \frac{1}{\sqrt{k!}}\,c_k\,y^k\,,
\quad y=(y_1,\dots,y_d),
\end{equation}
where $c_k$ are independent standard Gaussian random variables.
Observe that the random series in
(\ref{con4}) converges almost surely and it defines $g(y)$ as an entire
function. Let $g_1(y),\dots,g_p(y)$ be $p$ independent copies of
random function (\ref{con4}). Set
\begin{equation}\label{con5}
\vec g(y)=
\begin{pmatrix}
g_1(y) \\ \vdots \\ g_p(y))
\end{pmatrix},
\quad
h(y)=
\begin{pmatrix}
\frac{\partial g_1(y)}{\partial y_1} & \dots &
\frac{\partial g_1(y)}{\partial y_d} \\
\vdots & \ddots & \vdots \\
\frac{\partial g_p(y)}{\partial y_1} & \dots &
\frac{\partial g_p(y)}{\partial y_d}
\end{pmatrix}, 
\end{equation}
and
\begin{equation}\label{con6}
\| h(y)\|=\left[\sum_{k,j}
\left| \frac{\partial g_k(y)}{\partial y_j}\right|^2
\right]^{1/2}.
\end{equation}
Let $y^{(1)},\dots,y^{(m)}\in\R^d$, and 
let $D_m(\xi,\eta;y^{(1)},\ldots,y^{(m)})$ be the joint distribution
density of the Gaussian random tensors, 
$$
\xi=(\vec g(y^{(1)}),\dots,\vec g(y^{(m)})),\quad
\eta=(h(y^{(1)}),\ldots,h(y^{(m)})).
$$
Consider any point $\theta^0\in \RP^d$, different from
$P=(1,0,\dots,0)$. Consider any coordinate system in a neighborhood of
$\theta^0$, with the origin at $\theta^0$, 
 such that the metric tensor at $\theta^0$ is an identity tensor.
By $\theta^0+\frac{y}{\sqrt n}$, where $y\in \R^d$, we understand (for 
large $n$) a point in $\RP^d$ with coordinates $(\frac{y_1}{\sqrt
  n},\dots,\frac{y_d}{\sqrt n})$ in the chosen coordinate system.

\begin{theo}\label{univ_corr2} Assume that $\f(s)$, the characteristic
  function of $c_k$, satisfies estimates (\ref{pr2}), (\ref{pr2b})
of Theorem \ref{univ_corr}. 
Assume that the reference point $\theta^0\not= P\equiv
(1,0,\dots,0)\in\RP^d$ and $y^{(i)}\not=y^{(j)}$ for $i\not= j$. 
Then for every $m\ge 1$, there exists the limit, 
\begin{equation}\label{corn7}
\lim_{n\to\infty} K_{nm}(\theta^0+\frac{y^{(1)}}{\sqrt
  n},\ldots,\theta^0+\frac{y^{(m)}}{\sqrt n})= 
 K_m(y^{(1)},\ldots,y^{(m)})\,,
\end{equation}
where
\begin{equation}\label{con8}
K_m(y^{(1)},\ldots,y^{(m)})
=\int_{\R^{pd}} \cdots \int_{\R^{pd}}
\|\eta^{(1)}\|\cdots\|\eta^{(m)}\|
D_m(0,\eta;y^{(1)},\ldots,y^{(m)})d\eta^{(1)}\cdots d\eta^{(m)}. 
\end{equation} 
\end{theo}
  
{\bf Complex zeros of complex non-Gaussian multivariate random
  polynomials.}  There is a complex counterpart of Theorem
\ref{univ_corr2}. Consider multivariate polynomial (\ref{con1}) with
complex coefficients $c_k$ where $\{c_k\}$ are
independent identically distributed complex random variables such that 
\begin{equation}\label{con9}
\E c_k=0;\quad \E c_k^2=0,\quad \E |c_k|^2=1.
\end{equation}
Let $K_{nm}(z^{(1)},\dots,z^{(m)})$ be the $m$-point correlation
function of complex zeros of $f_n(z)$ in $\CP^d$, see 
\cite{BSZ1}-\cite{BSZ4} (for $d=1$ case see also earlier works
\cite{BBL1}, \cite{BBL2}, 
\cite{Ha}). By using our approach we are able to prove
the following theorem. Let us assume that the probability
distribution of 
$c_k$ is absolutely continuous with respect to the Lebesgue measure on 
the plane, and its characteristic function $\f(s)$ is infinitely
differentiable. As above, consider any point $\theta^0\in \CP^d$,
different from $P=(1,0,\dots,0)$, as a reference point.

\begin{theo}\label{univ_corr3} Assume that $\f(s)$, the characteristic
  function of $c_k$, satisfies the estimate,
\begin{equation}\label{con10}
|\f(s)|\le\frac{1}{(1+a s^2)^q}\,,
\end{equation}
 for some $a,q>0$,
and for any $j=(j_1,j_2)$ with $j_1+j_2\ge 2$ there exists
$c_j>0$ such that
\begin{equation}\label{con11}
\left|D^j\f(s)\right|<c_j\,,\quad
  D^j=\frac{\partial^{j_1+j_2}}{\partial s_1^{j_1}\partial
    s_2^{j_2}}\,. 
\end{equation}
Assume that the reference point $\theta^0\not= P\equiv
(1,0,\dots,0)\in\CP^d$ and $y^{(i)}\not=y^{(j)}$ for $i\not= j$. 
Then for every $m\ge 1$, there exists the limit, 
\begin{equation}\label{con12}
\lim_{n\to\infty} K_{nm}(\theta^0+\frac{y^{(1)}}{\sqrt
  n},\ldots,\theta^0+\frac{y^{(m)}}{\sqrt n})= 
 K_m(y^{(1)},\ldots,y^{(m)})\,,
\end{equation}
where
\begin{equation}\label{con13}
K_m(y^{(1)},\ldots,y^{(m)})
=\int_{\R^{pd}} \cdots \int_{\R^{pd}}
\|\eta^{(1)}\|^2\cdots\|\eta^{(m)}\|^2
D_m(0,\eta;y^{(1)},\ldots,y^{(m)})d\eta^{(1)}\cdots d\eta^{(m)}. 
\end{equation} 
\end{theo}

Theorems \ref{univ_corr2}, \ref{univ_corr3} can be further extended to 
non-Gaussian random sections of powers of line bundles over compact
manifolds (cf. 
\cite{BSZ1}-\cite{BSZ4}), but  we will not consider these extensions
here. 

\section{Appendix. Proof of Lemmas \ref{basic} and \ref{basic_add}}
\label{app}

{\it Proof of Lemma \ref{basic}}.
From (\ref{prove2}) and (\ref{pr2}) we have that
\begin{equation} \label{app1}
|\Phi_n(\ga)|\le
\prod_{k=0}^n \frac{1}{(1+a\om_k^2)^q}\,,
\end{equation}
This implies that
\begin{equation} \label{app1a}
|\Phi_n(\ga)|\le
\frac{1}{\left(1+a\sum_{k=0}^n\omega_k^2\right)^q}\,.
\end{equation}
To prove (\ref{pr3}) let us show that
\begin{equation} \label{app1b}
\sum_{k=0}^n \omega_k^2\ge C|\ga|^2\,.
\end{equation}
To do it observe that
\begin{equation} \label{app2}
\sum_{k=0}^n \omega_k^2=\sum_{i,j=1}^{2m} d_n(i,j)\ga_i\ga_j
\end{equation}
where the matrix $D_n=\left(d_n(i,j)\right)_{i,j=1}^{2m}$ is defined as
\begin{equation} \label{app3}
\begin{aligned}
D_n&=\begin{pmatrix}
A_n & B_n \\
B_n^T & C_n
\end{pmatrix},\\
A_n&=\left(a_n(s_i,s_j)\right)_{i,j=1}^m,
\quad B_n=\left(b_n(s_i,s_j)\right)_{i,j=1}^m,
\quad C_n=\left(c_n(s_i,s_j)\right)_{i,j=1}^m.
\end{aligned}
\end{equation}
By (\ref{corr11}),
\begin{equation} \label{app4}
\begin{aligned}
\lim_{n\to\infty}D_n&=D=\begin{pmatrix}
A & B \\
B^T & C
\end{pmatrix},\\
A&=\left(a(s_i,s_j)\right)_{i,j=1}^m,
\quad B=\left(b(s_i,s_j)\right)_{i,j=1}^m,
\quad C=\left(c(s_i,s_j)\right)_{i,j=1}^m.
\end{aligned}
\end{equation}
The matrix $D$ is positive definite (see \cite{BD}), hence
\begin{equation} \label{app5}
\sum_{i,j=1}^{2m} d(i,j)\ga_i\ga_j\ge C|\ga|^2\,.
\end{equation}
Due to (\ref{app4}), $D_n$ is also positive definite for large $n$,
uniformly in $n$.
This proves (\ref{app1b}).

Partition now all $k$'s into $T$ groups $M_j$ so that for each group
\begin{equation} \label{app6}
\sum_{k\in M_j} \om_k^2\ge \frac{C}{2T}\,|\ga|^2\,.
\end{equation}
Then 
\begin{equation} \label{app7}
|\Phi_n(\ga)|\le
\prod_{j=1}^T
\prod_{k\in M_j} \frac{1}{(1+a\om_k^2)^q}
\le \frac{1}{(1+a_0 |\ga|^2)^{Tq}}\,,
\quad a_0=\frac{C}{2T}\,a\,.
\end{equation}
Lemma \ref{basic} is proved. 

{\it Proof of Lemma \ref{basic_add}}. From (\ref{pr2b}) we obtain that
\begin{equation}\label{pr2a}
\left|\frac{d\f(s)}{ds}\right|<c_1|s|\,,
\end{equation}
where $c_1=c_2$. Consider first $k=(1,0,\ldots,0)$.
From (\ref{prove2})  we have that
\begin{equation} \label{add1}
\left|\frac{\partial\Phi_n(\ga)}{\partial \al_1}\right|=
\left|\sum_{k=0}^n \mu_k(s_1) \f'(\om_k)
\prod_{l\not= k}\f(\om_l)\right|\,.
\end{equation}
Observe that
\begin{equation} \label{add2}
\left|\prod_{l\not= k}\f(\om_l)\right|\le
\prod_{l\not= k} \frac{1}{(1+a\om_l^2)^q}
\le (1+|\ga|^2)\prod_{l=0}^n \frac{1}{(1+a\om_l^2)^q}\,,
\end{equation}
hence by (\ref{pr2a}) and Cauchy,
\begin{equation} \label{add3}
\left|\frac{\partial\Phi_n(\ga)}{\partial \al_1}\right|\le
c_1\left(\sum_{k=0}^n \mu_k^2(s_1)\right)^{1/2}
\left(\sum_{k=0}^n \om_k^2\right)^{1/2}
(1+|\ga|^2)\prod_{l=0}^n \frac{1}{(1+a\om_l^2)^q}\,.
\end{equation}
From (\ref{app2}),
\begin{equation} \label{add4}
\left(\sum_{k=0}^n \om_k^2\right)^{1/2}\le C|\ga|\,,
\end{equation}
hence
\begin{equation} \label{add5}
\left|\frac{\partial\Phi_n(\ga)}{\partial \al_1}\right|\le
C_0|\ga|
(1+|\ga|^2)\prod_{l=0}^n \frac{1}{(1+a\om_l^2)^q}\,.
\end{equation}
From (\ref{app7}) we obtain now that
\begin{equation} \label{add6}
\left|\frac{\partial\Phi_n(\ga)}{\partial \al_1}\right|\le
C_0|\ga|
(1+|\ga|^2)\frac{1}{(1+a_0 |\ga|^2)^{Tq}}\,,
\end{equation}
which implies that
\begin{equation} \label{add7}
\left|\frac{\partial\Phi_n(\ga)}{\partial \al_1}\right|\le
\frac{C_1}{(1+a_0 |\ga|^2)^{L}}\,,
\end{equation}
if we take $T$ sufficiently large. This proves estimate
(\ref{pr3a}) for $k=(1,0,\ldots,0)$. A similar argument
proves it for any $k$ with $|k|\equiv k_1+\ldots+k_{2m}=1$.
If $|k|\ge 2$, then formula (\ref{add1}) involves higher
order derivatives of $\f$ and different $s_j$'s. If the derivative
of $\f$ is of the second order or higher we get $\mu_k(s_j)$ 
or $\la_k(s_j)$ at least second power. If the derivative
of $\f$ is of the first order we get $\mu_k(s_j)\f'(\om_k)$
or $\la_k(s_j)\f'(\om_k)$. In the both cases we use (\ref{pr2a}),
(\ref{pr2b}) and (\ref{add4}) to prove that
\begin{equation} \label{add8}
\left|\frac{\partial^{|k|}\Phi_n(\ga)}{\partial \ga^k}\right|\le
C_0|\ga|^{|k|}
(1+|\ga|^2)^{|k|}\frac{1}{(1+a_0 |\ga|^2)^{Tq}}\,,
\end{equation}
which implies that
\begin{equation} \label{add9}
\left|\frac{\partial^{|k|}\Phi_n(\ga)}{\partial \ga^k}\right|\le
\frac{C_k}{(1+a_0 |\ga|^2)^{L}}\,,
\end{equation}
if we take $T$ sufficiently large. This proves estimate
(\ref{pr3a}) for any multi-index $k$. Lemma \ref{basic_add}
is proved.

\section*{Acknowledgment} The first author was supported in part by
NSF Grant DMS-9970625.

\end{document}